\documentclass{SciPost}

\usepackage{dsfont}
\usepackage{cite}
\usepackage{color}
\usepackage{tikz}
\usetikzlibrary{arrows,topaths,shapes.geometric,decorations.markings,shapes.misc,positioning}

\makeatletter
\newsavebox{\@brx}
\newcommand{\llangle}[1][]{\savebox{\@brx}{\(\m@th{#1\langle}\)}%
  \mathopen{\copy\@brx\kern-0.5\wd\@brx\usebox{\@brx}}}
\newcommand{\rrangle}[1][]{\savebox{\@brx}{\(\m@th{#1\rangle}\)}%
  \mathclose{\copy\@brx\kern-0.5\wd\@brx\usebox{\@brx}}}
\makeatother

\def\ii{{\rm i}}

\def\bra#1{\mathinner{\langle{#1}|}}
\def\ket#1{\mathinner{|{#1}\rangle}}
\def\dbra#1{\mathinner{\llangle{#1}|}}
\def\dket#1{\mathinner{|{#1}\rrangle}}

\newcommand{\scom}[2]{\big[#1,#2\big]}
\def\ua{\uparrow}
\def\da{\downarrow}
\newcommand{\stimes}{\circledast}
\def\lvac{{\bra{\rm vac}}}
\def\llvac{{\dbra{\rm vac}}}
\def\rvac{{\ket{\rm vac}}}
\def\rrvac{{\dket{\rm vac}}}
\def\g{{\textbf{g}}}
\def\u{{\textbf{u}}}
\def\bA{{\textbf{A}}}
\def\bB{{\textbf{B}}}
\def\bC{{\textbf{C}}}
\def\bD{{\textbf{D}}}
\def\bJ{{\textbf{J}}}
\def\bK{{\textbf{K}}}
\def\bL{{\textbf{L}}}
\def\bR{{\textbf{R}}}
\def\bS{{\textbf{S}}}

\def\bb{{\textbf{b}}}
\def\bh{{\textbf{h}}}

\def\bc{{\textbf{c}}}
\def\bn{{\textbf{n}}}
\def\bh{{\textbf{h}}}

\def\LL{{\mathbb{L}}}

\def\bxi{{\boldsymbol{\xi}}}
\def\bxid{{\overline{\boldsymbol{\xi}}}}

\def\hcalL{{\widehat{\mathcal{L}}}}

\def\calE{{\mathcal{E}}}

\begin{document}

\begin{center}{\Large \textbf{
Dissipation-driven integrable fermionic systems: from graded Yangians to exact nonequilibrium steady states
}}\end{center}

\begin{center}
Enej Ilievski\textsuperscript{1}, 
\end{center}

\begin{center}
{\bf 1} Institute for Theoretical Physics Amsterdam and Delta Institute for Theoretical Physics, University of Amsterdam, Science Park 904, 1098 XH Amsterdam, The Netherlands
\\
* e.ilievski@uva.nl
\end{center}

\begin{center}
\today
\end{center}


\section*{Abstract}
{\bf 
Using the Lindblad master equation approach, we investigate the structure of steady-state solutions of open integrable quantum
lattice models, driven far from equilibrium by incoherent particle reservoirs attached at the boundaries.
We identify a class of boundary dissipation processes which permits to derive exact steady-state density matrices
in the form of graded matrix-product operators. All the solutions factorize in terms of vacuum analogues of Baxter’s Q-operators
which are realized in terms of non-unitary representations of certain finite dimensional subalgebras of graded Yangians.
We present a unifying framework which allows to solve fermionic models and naturally incorporates higher-rank symmetries.
This enables to explain underlying algebraic content behind most of the previously-found solutions.
}

\vspace{10pt}
\noindent\rule{\textwidth}{1pt}
\tableofcontents\thispagestyle{fancy}
\noindent\rule{\textwidth}{1pt}
\vspace{10pt}

\section{Introduction}
\label{sec:intro}

Remarkable progress in experiments with cold 
atoms~\cite{Bloch_review,Kinoshita06,Cheneau12,Trotzky12,Gring12,Langen15,LGS16_review} has greatly impacted theoretical
research in the area of quantum many-body
dynamics~\cite{CC06,Polkovnikov_review,Eisert_review,Rigol_review,CC16_review,BD16_review,EF16_review,
IMPZ16_review,Cazalilla16_review}. Quantum systems which reside in the proximity of
a quantum integrable point have received a great amount of attention. Non-ergodic character of these systems was revealed through 
anomalous relaxation and absence of conventional thermalization, and paved the way to study new paradigms in quantum statistical 
mechanics such as pre-thermalization~\cite{MK10,KWE11,Marino12,SilvaPRL13,BertiniJSTAT15,BertiniPRL15} and equilibration
towards generalized Gibbs ensembles~\cite{RMO06,Rigol07,RSMS09,CEF11,CK12,CE13,DeNardis14,Wouters14,Pozsgay14,VR16_review}.
In an idealized scenario which neglects integrability-breaking perturbations, integrable interacting systems were shown to
permit a universal classification of local equilibria~\cite{IlievskiGGE15,StringCharge,IQC16} based on a complete set of local 
conservation laws~\cite{IMP15}.

Equilibrium statistical ensemble however constitute a fairly small set of quantum many-body states and are outside of perturbative 
regime insufficient to capture physically interesting situations in which systems support particle and energy currents.
An important step towards realizing accessing genuine far-from-equilibrium regimes is to devise an efficient computational
framework for accessing regimes of strongly-correlated quantum dynamics which often lie beyond the reach of traditional techniques. 
Switching from the Hamiltonian approach for closed systems to the \emph{open system} 
perspective~\cite{Gardiner_book,Breuer_book,Rivas_book} offers a promising route to achieve this.

A quantum system is regarded as an open system when as a result of interactions with its surroundings experiences incoherent loss of 
information (quantum decoherence), making a system evolving according to an effective \emph{non-unitary} evolution law.
In a highly controlled environment however, an irretrievable loss of information due to quantum noise may sometimes even act as
a resource~\cite{Diehl08,Diehl11,Barreiro11,Barontini13}. Quantum noise is typically modelled either as a stochastic process,
or alternatively via deterministic evolution laws in the form of quantum master equations, where the unitary dynamics is supplemented 
with additional non-Hamiltonian effective terms. The simplest master equations are Markovian~\cite{Lindblad76,GKS76} and thus entirely 
discard memory effects between a system and its environment.

Quantum master equations can be in many aspects perceived as quantum analogues of classical stochastic
models~\cite{KS68}. The latter encompassing a large class of systems which include asymmetric simple exclusion 
processes~\cite{Derrida92,Derrida93}, reaction-diffusion processes~\cite{Toussaint83,KRI,KRII,Alcaraz94}, zero-range 
processes~\cite{Spitzer91,Andjel84} and others (see e.g. \cite{Schutz} for a reivew).
While classical stochastic equations have been a subject of intense research in the past few decades which has lead to
many exactly solvable examples~\cite{Derrida92,Derrida93,SD93,Derrida98,Derrida01,deGier05,Blythe07,Gorissen12},
it is quite surprising that there exist merely a handful of recent theoretical studies of quantum master equation in the 
realm of low-dimensional many-particle systems~\cite{MarkoMBL,PZ09,Tomaz3Q,Bojan10,Marko10,Marko11,DK11,Marko15}.

Despite quantum many-body systems which undergo dissipation typically evolve to either trivial states, or highly entangled 
states of prohibitive complexity, there remarkably exist certain non-trivial examples of quantum dissipative Markovian dynamics
where an intricate interplay between noise and coherent evolution results in stationary state which are analytically tractable.
Integrability of the central model is of central importance here, as it makes it possible to identify Markovian particle reservoirs 
which, as explained in the manuscript, induce certain `symmetry protected' nonequilibrium states.
The main objective in this regard is to isolate scenarios in which the steady states of dissipative many-body dynamics
are of low complexity and permit an exact analytic description in terms of matrix-product states. This programme has been
initially employed in classical exclusion processes~\cite{Derrida92,Derrida93} and relatively recently
applied to quantum chain of non-interacting fermions~\cite{Znidaric10,ZnidaricPRE11}.
The same ideas have been shortly after expanded also to a few representative interacting exactly solvable many-body Hamiltonians,
such as the Heisenberg spin chain~\cite{ProsenPRL106,ProsenPRL107,KPS13} and the fermionic Hubbard 
model~\cite{Prosen_Hubbard,PP15}, which led to various applications (see e.g.~\cite{ProsenPRL106,PI13,Buca14,Marzolino14}).
For a historical perspective on the subject the reader is referred to the recent topical review \cite{Prosen_review}.

Despite many promising advancements on the subject, it is rather unsatisfactory that the structure of these solutions still remain
elusive and poorly understood. Indeed, no common framework which would explain the origin and meaning of integrable
dissipative boundaries and offer a systematic way to extend the results to more general scenarios has been proposed to the date. 
In particular, all previous attempts to understand the internal structure of these exactly solvable instances based
on `first symmetry principles' and algebraic concepts of Yang--Baxter integrability have been mainly unsuccessful, although
a few central insights have been made in \cite{IZ14,IP14} which unveiled the Lax formulation and highlighted the importance of
non-unitary representations of quantum groups. However, a comprehensive group-theoretic approach which would enable to
construct a larger class of solutions from first principles remained unknown.

The primary goal of this work is study formal aspects of integrable quantum chains driven far from equilibrium with aid of 
incoherent Markovian reservoirs attached to their ends. By continuing the bottom-up approach initiated in \cite{IZ14}, we shall
uncover the symmetry content behind some of the solutions obtained in the previous works, extend these results to models based
on higher rank algebras and discuss the paradigm of integrable steady states from the standpoint of representation 
theory of quantum algebras. This work offers a unifying algebraic construction for an entire class of exact 
nonequilibrium states belonging to the so-called rational integrable quantum spin chains by making use of
tools of quantum integrability theory. Specific instances which have been derived in the previous work with different 
techniques(cf.~\cite{ProsenPRL107,KPS13,IZ14,IP14}) are thus naturally incorporated in a common framework. Moreover, by using graded
vectors spaces and Lie superalgebras, we present how to accommodate for fermionic degrees of freedom~\cite{Kac77,Kac78}
and derive a new class of steady-state solutions for interacting integrable fermionic chains with $SU(n|m)$-symmetric Hamiltonians. 
Further quantitative analysis of the constructed solutions go beyond the main scope of the present study and will be thus omitted.
The task of computing correlation functions can however be carried out by using standard techniques based on matrix-product
states, see e.g.~\cite{Prosen_review}.

One of the central insights of our approach is rooted in the universal factorization property of quantum Lax operators.
This leads to the so-called `partonic' Lax operators can be regarded as the elementary constituents of Yang--Baxter 
integrable systems and are intrinsically related to the notion of
Baxter's Q-operators~\cite{Bazhanov10,Bazhanov11,Frassek11,Frassek13}, a widely used concept in the Bethe Ansatz diagonalization 
techniques. The observations that partonic Lax operators which realized over non-unitary irreducible modules may be used as local 
building units of exact steady-state solutions to certain Lindbladian dynamics is however a curious unconventional
feature which displays their proper nonequilibrium character.

\paragraph{Outline.}
The paper is organized as follows. In the preliminary section \ref{sec:prelim} we give a quick introduction to the
Lindblad master equation and briefly review some basic concepts regarding graded vectors spaces.
In section \ref{sec:Lindblad}, we proceed by introducing an out-of-equilibrium protocol by coupling a quantum chain of interacting
particles to incoherent particle reservoirs attached at its ends. We subsequently present the main algebraic structures which are 
afterwards used in the construction of the steady-state solutions. The notion of graded Yangians is defined in 
section~\ref{sec:Yangians}, and identify finite dimensional subalgebras which are intimately related to the Baxter's Q-operators.
In section \ref{sec:solutions} we outline a unifying construction for a class of non-trivial
current-carrying steady states, and present a few explicit examples of widely studied integrable spin and fermionic chains.
In section \ref{sec:vacuumQ} we provide some technical remarks on the notion
of vacuum Q-operators, and conclude in section \ref{sec:conclusion} by summarizing the main results and providing an outlook.

\section{Preliminaries}
\label{sec:prelim}

\subsection{Lindblad master equation}
In the approach of open quantum systems~\cite{Breuer_book,Rivas_book}, the time-evolution of a density operator $\rho(t)$
of the reduced density matrix of a central system (which presently represents a one-dimensional system of spins or interacting 
fermions) is governed by a completely positive and trace-preserving map $\mathcal{V}(t)$, reading compactly
\begin{equation}
\rho(t) = \mathcal{V}(t)\rho{(0)},\qquad \mathcal{V}(t) = \exp{(t\,\mathcal{L})}.
\label{eqn:Lindblad}
\end{equation}
The Liouville propagator obeys the semi-group property $\mathcal{V}(t_{1}+t_{2})=\mathcal{V}(t_{2})\mathcal{V}(t_{1})$.
Notice that, in contrast to the unitary propagator of the Hamiltonian evolution, the generator $\mathcal{V}(t)$ is not invertible.
The generator $\mathcal{L}$ takes the \emph{Lindblad form}~\cite{Lindblad76,GKS76}
\begin{equation}
\mathcal{L} = \mathcal{L}_{0} + \mathcal{D},
\label{eqn:Lindblad_split}
\end{equation}
where $\mathcal{L}_{0}\rho \equiv -\ii [H,\rho]$ is the ordinary Liouville--von Neumann unitary dynamics
generated by the Hamiltonian $H$ of the central system, while $\mathcal{D}$ is the dissipator which
fully encodes an \emph{effective} description of the environment and admits a canonical resolution in terms of
the Lindblad operators $A_{k}$,
\begin{equation}
\mathcal{D}\rho = \sum_{k}\Big(\big[A_{k},\rho A^{\dagger}_{k}\big] + \big[A_{k}\rho,A^{\dagger}_{k}\big]\Big).
\label{eqn:Lindblad_dissipator}
\end{equation}
Here each Lindblad `jump operator' $A_{k}$ acts as an independent incoherent process.\footnote{Lindbladian flows can be alternatively 
understood in terms of `quantum trajectories', i.e. an approach which uses a stochastic differential equation for an ensemble of pure 
quantum states evolving under an effective non-hermitian Hamiltonian~\cite{Dalibard92,Gardiner92,Plenio98}.} In our application, 
$A_{k}$ will be used to model incoming and outcoming particle flows through the boundaries of the quantum chain.
A particular advantange of such a nonequilibrium protocol is to have a simple setup for obtaining exact or approximate
results for genuine far-from-equilibrium states, reaching beyond the traditional linear response theory and quasi-stationary regimes
described with the hydrodynamic approach~\cite{BD16_review,Castro16,BCNF16}.

In this work we shall exclusively restrict our considerations to the \emph{steady states}. The latter correspond, by definition,
to fixed points of the Liouville dynamics, $\rho_{\infty}=\lim_{t\to \infty}\rho(t)$.
This means that a steady state is an operator $\rho_{\infty}$ from the \emph{kernel} (null space) of the generator $\mathcal{L}$,
\begin{equation}
\mathcal{L} \rho_{\infty} = 0.
\label{eqn:fixed_point}
\end{equation}
We will also encounter situations when $\dim \ker \mathcal{L}>1$, which physically corresponds to degenerate steady states
and leads to higher dimensional steady-state manifolds.

\subsection{Graded vector spaces}
In order to incorporate fermionic degrees of freedom in our description we shall make use of graded vector spaces.
A local Hilbert space attached to a site in a quantum chain is denoted by $\mathbb{C}^{n|m}$,
where integers $n$ and $m$ in the superscript signify the number of bosonic and fermionic states, respectively.
Below we briefly recall a few basic notions of graded vectors space and refer the reader for a more detailed exposition
to appendix \ref{app:SUSY}.

The two types of states are distinguished by
the $\mathbb{Z}_{2}$-parity,
\begin{equation}
p:\quad \{1,2,\ldots,n+m\} \to \{0,1\}.
\label{eqn:parity_mapping}
\end{equation}
The mapping $p$ equips $\mathbb{C}^{n+m}$ with a $\mathbb{Z}_{2}$-grading:
if $a$ belongs to a subset of bosonic (fermionic) indices we assign it a parity $p(a)=0$ ($p(a)=1$).
Gradation is naturally lifted to vectors spaces $\mathbb{C}^{n+m}$ and furthermore to the Lie algebra of
linear operators acting on $\mathbb{C}^{n+m}$. Specifically, by adopting the distinguished grading in which
$p(a)=0$ for $a\in \{1,2,\ldots n\}$ and $p(a)=1$ for $a\in \{n+1,\ldots m\}$, the space of $(n+m)$-dimensional matrices on 
$\mathbb{C}^{n+m}$ block-decompose into the bosonic (even) subspace $\mathcal{V}_{0}$ and fermionic (odd) subspace
$\mathcal{V}_{1}$. The two subspaces are typically referred to as the homogeneous components. The fundamental $\mathfrak{gl}(n|m)$ 
representation, denoted by $\mathcal{V}^{n|m}_{\square}$, is spanned by a basis of matrix units
$E^{ab}$, $(E^{ab})_{ij}=\delta_{a i}\,\delta_{j b}$. The action of the Lie bracket adjusted to the grading is expressed as
\begin{equation}
\Big[E^{ab},E^{cd}\Big] = \delta_{cb}\,E^{ad} - (-1)^{(a+b)(c+d)}\delta_{ad}\,E^{cb}.
\label{eqn:superspins}
\end{equation}
Since exchanging two fermionic states results in a minus sign, the presence of fermionic states in a graded tensor product
space non-trivially affects the multiplication rule. Namely, for a set of homogeneous elements\footnote{Homogeneous elements
are linear operators on $(\mathbb{C}^{n|m})^{\otimes N}$ with a well-defined parity.} we have
\begin{equation}
(A\otimes B)(C\otimes D) = (-1)^{BC}(AC\otimes BD).
\label{eqn:graded_TP}
\end{equation}
Tensor multiplication can be conveniently recast in the standard from by introducing the graded tensor product $\stimes$,
defined in accordance with $(A\stimes B)(C\stimes D) = AC\stimes BD$. Further clarifications about the notation can be
found in appendix \ref{app:SUSY}.

\section{Exactly solvable nonequilibrium steady states}
\label{sec:Lindblad}

The algebraic construction of the solutions which is outlined below consists two steps. The general strategy in some sense reminds
of solving a Poisson's equation. Namely, the first step is to identify a space of solutions for the bulk part which only accounts for 
the unitary part of the generator $\mathcal{L}_{0}$. Note that the entire space of bulk solutions is determined purely from the 
kinematic constraints, i.e. it is determined solely from the quantum symmetry algebra of the spin chain, irrespective of the 
representation labels.
The second step is to impose the dissipative boundary conditions which (when a solution exists) uniquely fixes the physical state
at hand. More specifically, this step amounts to chose suitable boundary auxiliary states and subsequently solve a non-linear
system of boundary constraints in the space of the free representation parameters. Such a separation of bulk and boundary processes is 
indeed a characteristic feature of all exactly solvable classical and quantum boundary-driven lattice models.

The aim of this section is to break down the entire procedure into elementary steps and systematically discuss all the necessary 
ingredients to carry out the algebraic construction for the class of steady-state solutions of integrable quantum chains.
The more difficult problem of identifying and classifying the relevant class of subalgebras is postponed to the next section, before 
finally presenting a few explicit results in Section \ref{sec:solutions}.

\subsection{Graded Yang--Baxter relation}
This work is focused on a particular class of integrable lattice models which involve both bosonic and 
fermionic states. These models can be systematically derived from the so-called rational solutions to the \emph{graded 
Yang--Baxter relation}. On a two-particle space $\mathbb{C}^{n|m}\otimes \mathbb{C}^{n|m}$ the latter takes the following form
\begin{equation}
R^{n|m}(z_{1}-z_{2})\big(\bL(z_{1})\stimes 1\big)\big(1 \stimes \bL(z_{2})\big) =
\big(1 \stimes \bL(z_{2})\big)\big(\bL(z_{1})\stimes 1\big)R^{n|m}(z_{1}-z_{2}),
\label{eqn:graded_YB}
\end{equation}
where $z_{1,2}$ are two arbitrary complex numbers usually referred to as the spectral parameters.
Here and subsequently we shall use the convention in which bold-faced symbols pertain to operators which act non-identically in the 
auxiliary space(s).

Let us first explain the main objects. The graded $R$-matrix $R^{n|m}(z)$ acts as an intertwiner on the two-fold space
$\mathbb{C}^{n|m}\otimes \mathbb{C}^{n|m}$, i.e. expresses the equivalence of two distinct orderings of the tensor product of
two $\bL$-operators. Matrices $R^{n|m}(z)$ are simply related to the graded permutation matrices $P^{n|m}$,
\begin{equation}
R^{n|m}(z)=z+P^{n|m},\qquad
P^{n|m} = (-1)^{b}E^{ab} \stimes E^{ba}= (-1)^{ab}E^{ab}\otimes E^{ba},
\label{eqn:graded_R}
\end{equation}
where matrix units $E^{ab}$ form the standard basis of linear operator in the fundamental module $\mathcal{V}^{n|m}_{\square}$.
The rational Yang--Baxter relation \eqref{eqn:graded_YB} can be formally understood as the \emph{defining relation} of an
infinite-dimensional associative algebra $\mathcal{Y}\equiv Y(\mathfrak{gl}(n|m))$ known as the \emph{Yangian}.
The $\bL$-operator from Eq.~\eqref{eqn:graded_YB} is in this context interpreted as a $\mathcal{Y}$-valued matrix on 
$\mathbb{C}^{n+m}$ which admits the resolution
\begin{equation}
\bL(z) = (-1)^{ab+b}E^{ab}\otimes \bL^{ab}(z).
\end{equation}
The class of solutions to the nonequilibrium protocol considered in this work turn out to be related to certain degenerate 
representation of $\mathcal{Y}$ which are defined discussed in Section \ref{sec:Yangians}.

\begin{figure}[t]
\centering
\begin{tikzpicture}[scale=0.6]
\tikzstyle{Lax} = [circle, thick, color=black, minimum width=16pt, fill=yellow!50, draw, inner sep=0pt]
\tikzstyle{R} = [circle, thick, color=black, minimum width=16pt, fill=violet!50, draw, inner sep=0pt]

\draw[<-,very thick,black] (0,2) node[left] {\scriptsize $3$} -- (5,-0.5) node[right] {\scriptsize $3$};
\draw[->,very thick,gray] (1,-2.75) node[below] {\scriptsize $1$} -- (5.25,2) node[above] {\scriptsize $1$};
\draw[->,very thick,gray] (3.25,-2.75) node[below] {\scriptsize $2$} -- (1,2.25) node[above] {\scriptsize $2$};
\node[Lax] (L1) at (1.5,1.25) {\scriptsize ${\bf L}$};
\node[Lax] (L2) at (3.65,0.15) {\scriptsize ${\bf L}$};
\node[R] (R1) at (2.5,-1) {\scriptsize $R$};

\path[-,thick,red] (6.25,-0.4) edge node[left] {} (6.75,-0.4);
\path[-,thick,red] (6.25,-0.6) edge node[left] {} (6.75,-0.6);


\draw[<-,very thick,black] (8,-0.5) node[left] {\scriptsize $3$} -- (13,-2.5) node[right] {\scriptsize $3$};
\draw[->,very thick,gray] (9.25,-2.75) node[below] {\scriptsize $1$} -- (11.75,2.25) node[above] {\scriptsize $1$};
\draw[->,very thick,gray] (12,-2.75) node[below] {\scriptsize $2$} -- (9.75,2.25) node[above] {\scriptsize $2$};
\node[Lax] (L3) at (11.75,-2) {\scriptsize ${\bf L}$};
\node[Lax] (L4) at (10,-1.25) {\scriptsize ${\bf L}$};
\node[R] (R2) at (10.7,0.1) {\scriptsize $R$};

\end{tikzpicture}
\caption{Graphical representation of the Yang--Baxter relation \eqref{eqn:graded_YB}. From the particle scattering perspective,
the relation imposes the equivalence of two apriori district ways of three consecutive pairwise elastic scatterings.
Time direction for physical particles flows vertically and is shown by gray trajectories.
The $R$-matrix $R(z_{1}-z_{2})$ acts proportionally to a graded permutation on two fundamental physical particles in a
$\mathfrak{su}(n|m)$ symmetric integrable quantum chain.
Lax operators ${\bf L}(z_{i}-z_{3})$ on the other hand govern scattering between physical particles with rapidities
$z_{i}$, $i\in \{1,2\}$, and a fictitious particle carrying rapidity $z_{3}$ whose time-direction runs horizontally.
Graded Yangians $Y(\mathfrak{gl}(n|m))$ are infinite-dimensional associative algebras
for the coefficients of the operator components of the ${\bf L}$-operator, with the Yang--Baxter equation
on $\mathbb{C}^{n|m}\otimes \mathbb{C}^{n|m}$ taking the role of its defining relations.}
\end{figure}
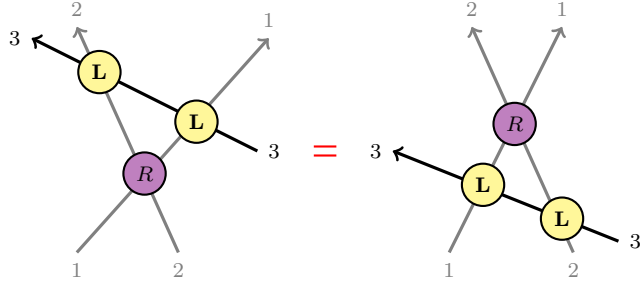

The presence of non-trivial grading can be seen as a diagonal `metric tensor' $\theta$ on two-particle space
$\mathbb{C}^{n+m}\otimes \mathbb{C}^{n+m}$,
\begin{equation}
\theta_{ac,bd}=(-1)^{ab}\delta_{ac}\delta_{bd},
\end{equation}
which allows for an alternative interpretation of Eq.~\eqref{eqn:graded_YB} as a braided Yang--Baxter equation on non-graded vector
space $\mathbb{C}^{n+m}\otimes \mathbb{C}^{n+m}$,
\begin{equation}
R^{n|m}(z_{1}-z_{2})\,\theta\,\big(\bL(z_{1})\otimes 1\big)\,\theta\,\big(1 \otimes \bL(z_{2})\big)=
\big(1 \otimes \bL(z_{2})\big)\,\theta\,\big(\bL(z_{1})\otimes 1\big)\,\theta\, R^{n|m}(z_{1}-z_{2}).
\end{equation}
The graded permutation can be expressed in terms of the non-graded permutation $P^{n+m}$ on $\mathbb{C}^{n+m}$ as
$P^{n|m}=\theta P^{n+m}$.

The central object of the algebraic Bethe Ansatz solution of integrable quantum models is the
fundamental transfer matrix $T^{n|m}_{\square}(z)$ operating on a $N$-particle physical space $(\mathbb{C}^{n|m})^{\otimes N}$
and satisfying the involution property (additional information can be found in appendix \ref{app:fusion})
\begin{equation}
T^{n|m}_{\square}(z) = {\rm Str}_{\mathcal{V}^{n|m}_{\square}}\;
\bL_{\square}(z)\stimes \cdots \stimes \bL_{\square}(z),\qquad \left[T^{n|m}_{\square}(z),T^{n|m}_{\square}(z^{\prime})\right]=0.
\end{equation}
Commutativity of transfer matrices is ensured by the existence of the $R$-matrices $R^{n|m}_{\square,\square}(z)$ which
intertwines two fundamental \emph{auxiliary} representations $\mathcal{V}^{n|m}_{\square}$, i.e. it solves the corresponding
(graded) Yang--Baxter relation. We need to emphasize however that Eq.~\eqref{eqn:graded_YB} is written with the opposite
identification of physical and auxiliary degrees of freedom with respect to the form which is most commonly used in the (algebraic) 
Bethe ansatz technique. The upshot is that our construction necessitates generic auxiliary models and not the conventional
fundamental auxiliary representations. Indeed, the ordinary set of transfer matrices $T^{n|m}_{\square}(z)$ and their fused 
counterparts which correspond to finite-dimensional auxiliary irreducible representations of $\mathfrak{gl}(n|m)$ are \emph{unitary} 
objects and in fact have no natural place in our application.

\paragraph{Differential Yang--Baxter relation.}
Taking the derivative of Eq.~\eqref{eqn:graded_YB} with respect to $z=z_{1}-z_{2}$ yields
the \emph{differential Yang--Baxter relation} (sometimes also called the Sutherland relation, 
cf.~\cite{Sutherland70,Thacker86,KPS13,IZ14}),
\begin{equation}
\big[h^{n|m},\bL(z) \stimes \bL(z)\big] = \bL(z)\stimes \bL^{\prime}(z) - \bL^{\prime}(z)\stimes \bL(z),
\label{eqn:Sutherland}
\end{equation}
using the short-handed notation $\bL^{\prime}(z)\equiv \partial_{z}\bL(z)$.
Equation \eqref{eqn:Sutherland} is simply a consequence of the fact that $\mathfrak{su}(n|m)$-symmetric Hamiltonian densities
$h^{n|m}$ coincide with graded permutations $P^{n|m}$ over $\mathbb{C}^{n|m}\otimes \mathbb{C}^{n|m}$, i.e.
\begin{equation}
h^{n|m} = P^{n|m}\partial_{z} R^{n|m}(z) = P^{n|m}.
\label{eqn:hamiltonian_density}
\end{equation}
For the so-called rational spin chains, relation \eqref{eqn:Sutherland} is in fact a simple corollary the zero-curvature property of 
the Lax connection.\footnote{Relation \eqref{eqn:Sutherland} should not be confused with the lattice version of the Lax representation 
which takes the local form
$\partial_{t}\bL_{i}(z)=\ii[H,\bL_{i}(z)]=\mathbf{A}_{i+1}(z)\bL_{i}(z)-\bL_{i}(z)\mathbf{A}_{i}(z)$, with matrices 
$\bL_{i}(z)$ and $\mathbf{A}_{i}(z)$ corresponding to the spatial and temporal component of the (discrete) connection of the
associated to the auxiliary linear problem.}
What is more important is that the differential Yang--Baxter relation \eqref{eqn:Sutherland_general} is satisfied on
a purely \emph{algebraic} level, i.e. irrespective of a representations of the auxiliary components of the $\bL$-operator.
A general solution to Eq.~\eqref{eqn:hamiltonian_density} is given by an operator $\bL_{\Lambda_{n+m}}(z)$ acting on
a product space of a local physical space and an arbitrary auxiliary representations, that is 
$\mathcal{V}_{\square}\otimes \mathcal{V}^{+}_{\Lambda_{n+m}}$. Here $\mathcal{V}^{+}_{\Lambda_{n+m}}$ denotes a
generic \emph{irreducible highest-weight} representation of $\mathfrak{gl}(n|m)$ Lie superalgebra, characterized by a set of Dynkin 
labels $\Lambda_{n+m}$ (cf. appendix \ref{app:fusion}).

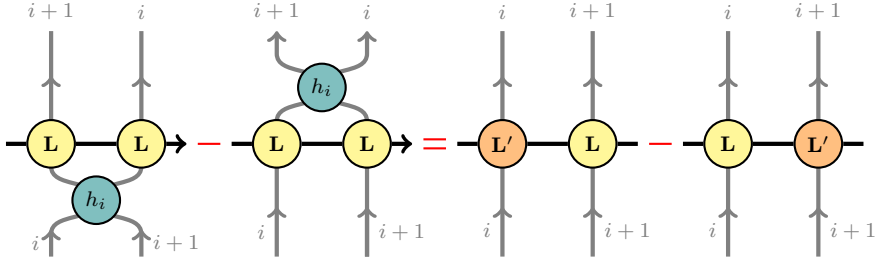
\begin{figure}[h]
\label{fig:Sutherland}
\centering
\begin{tikzpicture}[scale=0.6]

\tikzstyle{interaction} = [circle, thick, color=black, minimum width=18pt, fill=teal!50, draw, inner sep=0pt]
\tikzstyle{Lax} = [circle, thick, color=black, minimum width=18pt, fill=yellow!50, draw, inner sep=0pt]
\tikzstyle{dLax} = [circle, thick, color=black, minimum width=18pt, fill=orange!50, draw, inner sep=0pt]

\path[<-, ultra thick,gray] (0,3.5) edge node[left] {} (0,2.5);
\path[-, ultra thick,gray] (0,4.5) edge node[left] {} (0,3.4) node[above] {\scriptsize ${i+1}$};
\path[<-, ultra thick,gray] (2,3.5) edge node[left] {} (2,2.5);
\path[-, ultra thick,gray] (2,4.5) edge node[left] {} (2,3.4) node[above] {\scriptsize $i$};
\draw [-, ultra thick,gray] (0,2.5) -- (0,1.5) to[out=-90,in=90] (1,0.75) to[out=-90,in=90] (0,0);
\draw [-, ultra thick,gray] (2,2.5) -- (2,1.5) to[out=-90,in=90] (1,0.75) to[out=-90,in=90] (2,0);
\path[<-, ultra thick,gray] (0,0.1) edge node[left] {\scriptsize $i$} (0,-0.5);
\path[<-, ultra thick,gray] (2,0.1) edge node[right] {\scriptsize $i+1$} (2,-0.5);
\node[interaction] (h) at (1,0.75) {\scriptsize $h_{i}$};
\node[Lax] (L1) at (0,2) {\scriptsize ${\bf L}$};
\node[Lax] (L2) at (2,2) {\scriptsize ${\bf L}$};
\path[-, ultra thick, color=black] (-1,2) edge node[right] {} (L1)
			(L1) edge node[left] {} (L2);
\path[->,ultra thick, color=black] (L2) edge node[left] {} (3,2);

\path[-, thick, color=red] (3.25,2) edge node[left] {} (3.75,2);

\path[->, ultra thick,gray] (5,-0.5) edge node[left] {\scriptsize $i$} (5,0.6);
\path[->, ultra thick,gray] (7,-0.5) edge node[right] {\scriptsize $i+1$} (7,0.6);
\draw [->, ultra thick,gray] (5,0.5) -- (5,2.5) to[out=90,in=-90] (6,3.25) to[out=90,in=-90] (5,4) -- (5,4.5) node[above] {\scriptsize ${i+1}$};
\draw [->, ultra thick,gray] (7,-0.5) -- (7,2.5) to[out=90,in=-90] (6,3.25) to[out=90,in=-90] (7,4) -- (7,4.5) node[above] {\scriptsize $i$};
\node[interaction] (h) at (6,3.25) {\scriptsize $h_{i}$};
\node[Lax] (L3) at (5,2) {\scriptsize ${\bf L}$};
\node[Lax] (L4) at (7,2) {\scriptsize ${\bf L}$};
\path[-, ultra thick, color=black] (4,2) edge node[right] {} (L3)
			(L3) edge node[left] {} (L4);
\path[->, ultra thick, color=black] (L4) edge node[left] {} (8,2);

\path[-, thick, color=red] (8.25,2.1) edge node[left] {} (8.75,2.1);
\path[-, thick, color=red] (8.25,1.9) edge node[left] {} (8.75,1.9);

\draw [->,ultra thick,gray] (10,-0.5) -- (10,0.6);
\draw [->,ultra thick,gray] (10,0.5) node[below left] {\scriptsize $i$} -- (10,3.5);
\draw [-,ultra thick,gray] (10,3.4) -- (10,4.5) node[above] {\scriptsize $i$};
\draw [->,ultra thick,gray] (12,-0.5) -- (12,0.6);
\draw [->,ultra thick,gray] (12,0.5) node[below right] {\scriptsize $i+1$} -- (12,3.5);
\draw [-,ultra thick,gray] (12,3.4) -- (12,4.5) node[above] {\scriptsize $i+1$};
\node[dLax] (L5) at (10,2) {\scriptsize ${\bf L^{\prime}}$};
\node[Lax] (L6) at (12,2) {\scriptsize ${\bf L}$};
\path[-, ultra thick, color=black] (9,2) edge node[right] {} (L5)
			(L5) edge node[left] {} (L6)
			(L6) edge node[left] {} (13,2);

\path[-, thick, color=red] (13.25,2) edge node[left] {} (13.75,2);

\draw [->,ultra thick,gray] (15,-0.5) -- (15,0.6);
\draw [->,ultra thick,gray] (15,0.5) node[below left] {\scriptsize $i$} -- (15,3.5);
\draw [-,ultra thick,gray] (15,3.4) -- (15,4.5) node[above] {\scriptsize $i$};
\draw [->,ultra thick,gray] (17,-0.5) -- (17,0.6);
\draw [->,ultra thick,gray] (17,0.5) node[below right] {\scriptsize $i+1$} -- (17,3.5);
\draw [-,ultra thick,gray] (17,3.4) -- (17,4.5) node[above] {\scriptsize $i+1$};
\node[Lax] (L7) at (15,2) {\scriptsize ${\bf L}$};
\node[dLax] (L8) at (17,2) {\scriptsize ${\bf L^{\prime}}$};
\path[-, ultra thick, color=black] (14,2) edge node[right] {} (L7)
			(L7) edge node[left] {} (L8)
			(L8) edge node[left] {} (18,2);

\end{tikzpicture}
\caption{The differential Yang--Baxter equation (see Eq.~\eqref{eqn:Sutherland}) takes the form of an
operator-valued divergence condition on a one-dimensional lattice. The left-hand side is a schematic representation of
the local action of the unitary propagator $\partial_{t}{\Omega_{N}}\simeq [H,\Omega_{N}]$ which produces a telescopic sum of
terms with a single `defect operator' which coincides with the derivative of the ${\bf L}$-operator (shown in orange).}
\end{figure}

A key property of algebraic relation \eqref{eqn:Sutherland} is that it remains intact under fusion of auxiliary spaces.
This readily makes it possible to extend it to composite (many-particle) auxiliary spaces, namely
we may quite generally consider a multi-component Lax operators of the following form
\begin{equation}
\LL_{\boldsymbol{\Lambda}}({\bf z}) \equiv \bL_{\Lambda^{1}_{n+m}}(z_{1})\otimes \bL_{\Lambda^{2}_{n+m}}(z_{2})\otimes 
\cdots \otimes \bL_{\Lambda^{\ell}_{n+m}}(z_{\ell}),
\label{eqn:Sutherland_general}
\end{equation}
acting on $\mathcal{V}_{\square} \otimes \mathcal{H}_{\rm aux}$, with $\mathcal{H}_{\rm aux}$ representing
an arbitrary $\ell$-component auxiliary product space
$\mathcal{H}_{\rm aux}\cong\mathcal{V}_{\Lambda^{1}_{n+m}} \otimes \cdots \otimes \mathcal{V}_{\Lambda^{\ell}_{n+m}}$ characterized
by a set of weight vectors $\boldsymbol{\Lambda}\equiv \{\Lambda^{1}_{n+m},\ldots,\Lambda^{\ell}_{n+m}\}$ and a vector of
complex parameters ${\bf z}\equiv \{z_{1},\ldots, z_{\ell}\}$.
It is worthwhile emphasizing at this point that the tensor product in Eq.~\eqref{eqn:Sutherland_general} is written with respect
to auxiliary spaces $\mathcal{V}_{\Lambda_{n+m}}$, and thus differs from the tensor product of two Lax operator from 
Eq.~\eqref{eqn:graded_YB} which multiplies two copies of local physical (fundamental) spaces $\mathbb{C}^{n|m}$.
The multi-component Lax operator $\LL_{\boldsymbol{\Lambda}}({\bf z})$ obeys an analogue of Eq.~\eqref{eqn:Sutherland},
where the $z$-derivative acting on $\bL_{\Lambda_{n+m}}(z)$ should be replaced by the chain-rule derivation
$\partial_{{\bf z}}\equiv \sum_{i=1}^{\ell}\partial_{z_{i}}$ on $\LL_{\boldsymbol{\Lambda}}({\bf z})$, as
illustrated in Figure~\ref{fig:two-leg}.

\subsection{Amplitude factorization}
We consider an $N$-site quantum system with the Hamiltonian $H^{n|m}=\sum_{i=1}^{N-1}h^{n|m}_{i}$ and
impose open boundary conditions. This class of models describes integrable quantum chains symmetric under $\mathfrak{su}(n|m)$ Lie 
superalgebra~\cite{Kulish86,KS90} whose interactions take a simple form
\footnote{Here and throughout the text we afforded an unambiguous abuse of notation and replaced all parities $p(a)$ with
in the superscripts by their argument $a$, i.e. wrote simply $(-1)^{p(a)}\to (-1)^{a}$.}
\begin{equation}
h^{n|m} = (-1)^{b}E^{ab}\stimes E^{ba}.
\label{eqn:integrable_interaction}
\end{equation}
We adopt the convention for summing over repeated indices throughout the text, unless stated otherwise.

In the boundary-driven setting, the Lindblad dissipator $\mathcal{D}$ gets naturally split into two \emph{independent}
incoherent processes assigned to the boundaries of the chain, i.e. $\mathcal{D}=\mathcal{D}_{\rm L}+\mathcal{D}_{\rm R}$, where
$\mathcal{D}_{\rm L}$ ($\mathcal{D}_{\rm R}$) operates only on the first (last) site of the chain.
It is perhaps not too surprising that fixed-point solutions $\rho_{\infty}$ to Eq.~\eqref{eqn:fixed_point} for some bulk Hamiltonian 
$H^{n|m}$ with \emph{generic} Lindblad boundary dissipators typically yields density matrices lacking any obvious structure. 
Remarkably however, there exist a classes of dissipative boundary conditions for which one may derive an exact algebraic expression 
for it. Before presenting the precise form of such integrable dissipative boundaries in section \ref{sec:boundary_conditions},
we first wish to explain why localizing the dissipators to the chain boundaries plays a vital role in our construction and
briefly comment on some important consequences. In simple terms, attaching the dissipators only to the boundaries manifestly ensures 
that the unitary part of the fixed-point condition \eqref{eqn:fixed_point} annihilates $\rho_{\infty}$, modulo some residual
terms which stick at the boundary sites of the chain. This neat property motivates to use the algebra of (possibly non-local) 
commuting operators associated to the Hamiltonian $H^{n|m}$ as a trial space of operators for constructing an appropriate
$\Omega$-amplitude introduced in Eq.~\eqref{eqn:amplitude_factrorization}. In other words, by assuming that the steady-state solution 
of our problem has a well-defined local structure which is related to the symmetry algebra of the Hamiltonian. As explained below, 
the global symmetry gets broken only due to a mismatch in the boundary conditions, which is essentially
the reason why the steady state is non-trivial, $[H,\rho_{\infty}]\neq 0$.

We now proceed by employing the following \emph{amplitude factorization} of the density operator $\rho_{\infty}$,
\begin{equation}
\rho_{\infty} = \Omega_{N}\,\Omega_{N}^{\dagger}.
\label{eqn:amplitude_factrorization}
\end{equation}
Let us immediately stress that even though such a decomposition can be applied quite generally, it plays no fundamental
role without imposing further restrictions on the amplitude operators $\Omega_{N}$.\footnote{Notice that, for instance, the 
factorization property \eqref{eqn:amplitude_factrorization} is not gauge-invariant as it notably exhibits a unitary freedom of square 
roots, i.e. $\Omega \to \Omega U$ for some unitary matrix $U$.}. Indeed, the factorization property has been originally observed 
already in the seminal paper \cite{ProsenPRL107} where a non-pertrubative steady-state solution of the driven anisotropic, where it is 
referred to as the`reverse many-body Cholesky factorization'. However, in the class of solutions considered here, $\Omega_{N}$ need 
not be a Cholesky factor of a steady state $\rho_{\infty}$, namely there is no requirement that $\Omega_{N}$ takes a triangular form 
when expanded in the standard many-body computational basis of unit matrices spanning $(\mathbb{C}^{m+n})^{\otimes N}$. Nonetheless, 
since the entire class of solutions which are presented below extends the simplest $\mathfrak{su}(2)$ model to higher dimensional 
quantum spaces, we shall adopt the factorization property as a starting point of our presentation\footnote{While the factorization 
property can be sometimes inferred by inspecting the structure of exact solution found by symbolic algebra routines for small enough 
instances, its origin and physical significance remains elusive at the moment.}.

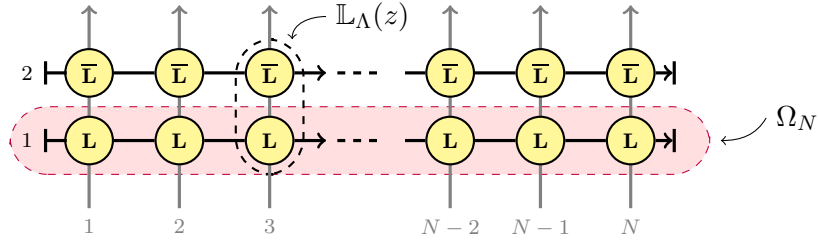
\begin{figure}[h]
\centering
\begin{tikzpicture}[scale=0.6]
\tikzstyle{Lax} = [circle, thick, color=black, minimum width=18pt, fill=yellow!50, draw, inner sep=0pt]
\tikzstyle{dLax} = [circle, thick, color=black, minimum width=18pt, fill=orange!50, draw, inner sep=0pt]

\draw [rounded corners=5mm, fill=pink!50, dashed, draw=purple] (-1.75,-0.75)--(13.75,-0.75)--(13.75,0.75)--(-1.75,0.75)--cycle;

\draw (15,0.5) edge[out=-120,in=0,->] (14,0) node[right] {$\Omega_{N}$};

\draw [rounded corners=5mm, fill=none, thick, dashed, draw=black] (3.25,-0.75)--(4.75,-0.75)--(4.75,2.25)--(3.25,2.25)--cycle;

\draw (5.25,2.75) edge[out=-180,in=90,->] (4.5,2.25) node[right] {$\mathbb{L}_{\Lambda}(z)$};

\draw [->,very thick,gray] (0,-1.5) node[below] {\scriptsize $1$} -- (0,3);
\draw [->,very thick,gray] (2,-1.5) node[below] {\scriptsize $2$} -- (2,3);
\draw [->,very thick,gray] (4,-1.5) node[below] {\scriptsize $3$} -- (4,3);

\draw [->,very thick,gray] (8,-1.5) node[below] {\scriptsize $N-2$} -- (8,3);
\draw [->,very thick,gray] (10,-1.5) node[below] {\scriptsize $N-1$}-- (10,3);
\draw [->,very thick,gray] (12,-1.5) node[below] {\scriptsize $N$} -- (12,3);

\node[Lax] (L1) at (0,0) {\scriptsize ${\bf L}$};
\node[Lax] (L2) at (2,0) {\scriptsize ${\bf L}$};
\node[Lax] (L3) at (4,0) {\scriptsize ${\bf L}$};
\draw[|-,very thick,black] (-1,0) node[left] {\scriptsize $1$} -- (-0.5,0);
\path[-, very thick,black] (-0.5,0) edge node[right] {} (L1)
			(L1) edge node[left] {} (L2)
			(L2) edge node[left] {} (L3);
\draw[->,very thick,black] (L3) edge node[left] {} (5.25,0);

\draw[dashed,very thick,black] (5.5,0) edge node[left] {} (6.5,0);

\node[Lax] (L4) at (8,0) {\scriptsize ${\bf L}$};
\node[Lax] (L5) at (10,0) {\scriptsize ${\bf L}$};
\node[Lax] (L6) at (12,0) {\scriptsize ${\bf L}$};
\path[-, very thick,black] (7,0) edge node[right] {} (L4)
			(L4) edge node[left] {} (L5)
			(L5) edge node[left] {} (L6);
\draw[->|,very thick,black]	(L6) edge node[left] {} (13,0);

\node[Lax] (U1) at (0,1.5) {\scriptsize ${\bf \overline{L}}$};
\node[Lax] (U2) at (2,1.5) {\scriptsize ${\bf \overline{L}}$};
\node[Lax] (U3) at (4,1.5) {\scriptsize ${\bf \overline{L}}$};
\draw[|-,very thick,black] (-1,1.5) node[left] {\scriptsize $2$} -- (-0.5,1.5);
\path[-,very thick,black] (-0.5,1.5) edge node[right] {} (U1)
			(U1) edge node[left] {} (U2)
			(U2) edge node[left] {} (U3);
\draw[->,very thick,black]	(U3) edge node[left] {} (5.25,1.5);

\draw[dashed,very thick,black] (5.5,1.5) edge node[left] {} (6.5,1.5);

\node[Lax] (U4) at (8,1.5) {\scriptsize ${\bf \overline{L}}$};
\node[Lax] (U5) at (10,1.5) {\scriptsize ${\bf \overline{L}}$};
\node[Lax] (U6) at (12,1.5) {\scriptsize ${\bf \overline{L}}$};
\path[-, very thick, black] (7,1.5) edge node[right] {} (U4)
			(U4) edge node[left] {} (U5)
			(U5) edge node[left] {} (U6);
\draw[->|,very thick,black] (U6) edge node[left] {} (13,1.5);

\end{tikzpicture}
\caption{Matrix-product representation of the non-equilibrium steady state $\rho_{\infty}=\Omega_{N}\Omega^{\dagger}_{N}$:
the amplitude operator $\Omega_{N}$ is represented by the degrees of freedom residing in the bottom row (shown in pink), while
its conjugate transpose $\Omega^{\dagger}_{N}$ corresponds to the upper row.
In terms of an auxiliary scattering process, auxiliary particles are depicted by black lines and propagate in the horizontal
direction. They can be viewed as fictitious particles composed of canonical bosons, fermions
of complex (super)spins, emanating from the auxiliary vacuum on one end and getting absorbed by the same vacuum at the other end.
Physical degrees of freedom (shown by gray vertical lines) are on the other hand associated to $N$ fundamental particles of 
$\mathfrak{gl}(n|m)$ Lie superalgebra. The off-shell steady-state density operator admits an interpretation as a vacuum contraction 
of a homogeneous two-row monodromy operator
$\mathbb{M}_{\Lambda}(z) = \mathbb{L}_{\Lambda}(z)\stimes \mathbb{L}_{\Lambda}(z)\stimes \cdots \stimes \mathbb{L}_{\Lambda}(z)$,
where $\mathbb{L}_{\Lambda}(z)=\bL_{\Lambda}(z)\otimes \bar{\bL}_{\Lambda}(z)$ is a Lax operator which acts on
a vertical rung.}
\label{fig:MPS}
\end{figure}

Following the above reasoning, the local symmetry of model is manifestly realized by introduce the following
homogeneous \emph{fermionic matrix-product operator}
\begin{equation}
\Omega_{N}({\bf g}) = \lvac \bL({\bf g})\stimes \bL({\bf g})\stimes \cdots \stimes \bL({\bf g})\rvac,
\label{eqn:Omega_abstract}
\end{equation}
acting on an $N$-site quantum chain, with symbol $\stimes$ designating the graded tensor product and takes into account the presence 
of fermionic states. In the pictorial representation, the amplitude represents the lower leg
in Figure \ref{fig:MPS}. The key properties of the amplitude operator are:
\begin{itemize}
\item Each tensor factor in Eq.~\eqref{eqn:Omega_abstract} is assumed to be a $\mathfrak{gl}(n|m)$-invariant
Lax operator parametrized by a continuous real parameter $\g$ (being the coupling strength parameter associated the
Lindblad dissipator $\mathcal{D}$).
The $\bL$-operator acts (by definition) on a local physical space $\mathbb{C}^{n|m}$ and an auxiliary Hilbert space which
is at the moment left unspecified. Hence, the auxiliary component of the $\bL$-operator can be thought of as
a generic representation of the underlying quantum algebra.
\item We have introduced the boundary state $\rvac$ which will be subsequently referred to as the (auxiliary) \emph{vacuum}.
The vacuum state is determined by the choice integrable dissipative boundaries. In all the instances addressed in this work,
$\rvac$ is simply an `empty state', i.e. a product of highest- or lowest-weight vectors from the irreducible components which form a 
representation of the auxiliary algebra of the $\bL$-operator. This uniquely fixes the vacuum state once the representation
labels (i.e. Dynkin labels and additional labels to specify the types of modules involved) associated to the $\bL$-operator
from Eq.~\eqref{eqn:Omega_abstract} are being specified. The role of the auxiliary vacua shall be more carefully explained
in Section \ref{sec:solutions} we treat a few explicit instances.
\end{itemize}

Let us now return to the differential Yang--Baxter relation.
Algebraic property \eqref{eqn:Sutherland} can be readily extended to the entire spin chain Hilbert space
$\mathcal{H}\cong (\mathbb{C}^{n+m})^{\otimes N}$ by simply expanding out the commutator $[H^{n|m},\Omega_{N}]$
and iteratively applying Eq.~\eqref{eqn:Sutherland} at every pair of adjacent lattice sites.
This results in a telescoping cancellation mechanism which on globally almost annihilates the unitary part of the evolution
generated by $\mathcal{L}_{0}$, leaving behind only residual boundary terms which are an artefact of open boundary condition.
This can be formally expressed in the form~\cite{ZnidaricJPA,KPS13}
\begin{equation}
\big[H^{n|m},\Omega_{N}(\g)\big] = \Xi_{\rm L}\otimes \Omega_{N-1}(\g) - \Omega_{N-1}(\g)\otimes \Xi_{\rm R},
\label{eqn:divergence}
\end{equation}
which is can be viewed as the global version of the local condition \eqref{eqn:Sutherland} after contracting with the
vacuum $\rvac$ at the end.
We have written $\Xi_{\rm L,R}$ to denote a pair of `boundary defect operators' acting only in the boundary particle spaces
(their explicit form is not of our interest).

Factorization property \eqref{eqn:amplitude_factrorization} indicates that the auxiliary Hilbert space
$\mathcal{H}_{\rm aux}$ associated to the matrix-product representation of the steady-state density matrix $\rho_{\infty}$
is a two-fold product of auxiliary subspaces which belong to mutually conjugate realizations of the underlying symmetry.
Therefore, setting $\ell=2$ in Eq.~\eqref{eqn:Sutherland_general} and writing shortly $\Lambda_{m+n}\equiv \Lambda$, we arrive at
the `off-shell' representation\footnote{An `off-shell' operator is referred to an object of an appropriate algebraic form which is
\emph{not required} to be a solution of the fixed-point condition \eqref{eqn:fixed_point}.} for
$\rho_{\boldsymbol{\Lambda}}({\bf z})$,
\begin{equation}
\rho_{\Lambda}(z) = \llvac \LL_{\Lambda}(z) \stimes \LL_{\Lambda}(z) \stimes \cdots \LL_{\Lambda}(z) \rrvac,
\label{eqn:ss_universal}
\end{equation}
where
\begin{equation}
\LL_{\Lambda}(z) = \bL_{\Lambda}(z)\otimes \overline{\bL}_{\Lambda}(z).
\label{eqn:double_L}
\end{equation}
is a two-row Lax operator of the form which is represented in Figure~\ref{fig:MPS} by a vertical rung.
Similarly, the boundary state $\rrvac$ represents a factorizable state of two auxiliary vacua, $\rrvac= \ket{0} \otimes \ket{0}$.
The internal structure of the vacuum state $\rvac$, which depends on the rank of symmetry algebra and the choice of integrable 
boundaries, will be detailed out in Section \ref{sec:solutions}.

\begin{figure}[h]
\label{fig:two-leg}
\centering
\begin{tikzpicture}[scale=0.3]
\tikzstyle{interaction} = [circle, thick, color=black, minimum width=10pt, fill=teal!50, draw, inner sep=0pt]
\tikzstyle{Lax} = [circle, thick, color=black, minimum width=10pt, fill=yellow!50, draw, inner sep=0pt]
\tikzstyle{dLax} = [circle, thick, color=black, minimum width=10pt, fill=orange!50, draw, inner sep=0pt]

\path[-,very thick,gray] (0,2) edge node[left] {} (0,7);
\path[-,very thick,gray] (2,2) edge node[left] {} (2,7);
\draw [-,very thick,gray] (0,2.5) -- (0,1.5) to[out=-90,in=90] (1,0.75) to[out=-90,in=90] (0,0);
\draw [-,very thick,gray] (2,2.5) -- (2,1.5) to[out=-90,in=90] (1,0.75) to[out=-90,in=90] (2,0);
\path[<-,very thick,gray] (0,0.1) -- (0,-0.5);
\path[<-,very thick,gray] (2,0.1) -- (2,-0.5);
\node[interaction] (h) at (1,0.75) {};
\node[Lax] (L1) at (0,2) {};
\node[Lax] (L2) at (2,2) {};
\node[Lax] (L3) at (0,5) {};
\node[Lax] (L4) at (2,5) {};
\path[-,very thick, color=black] (-1,2) edge node[right] {} (L1)
			(L1) edge node[left] {} (L2);
\path[-,very thick, color=black] (L2) edge node[left] {} (3,2);
\path[-, very thick, color=black] (-1,5) edge node[right] {} (L3)
			(L3) edge node[left] {} (L4);
\path[-,very thick, color=black] (L4) edge node[left] {} (3,5);

\path[-,thick, color=red] (3.25,3.6) edge node[left] {} (3.75,3.6);
\path[-,thick, color=red] (3.25,3.4) edge node[left] {} (3.75,3.4);


\path[-,very thick,gray] (5,0) edge node[left] {} (5,2);
\path[-,very thick,gray] (5,5) edge node[left] {} (5,7);
\path[-,very thick,gray] (7,0) edge node[left] {} (7,2);
\path[-,very thick,gray] (7,5) edge node[left] {} (7,7);

\draw [-,very thick,gray] (5,5.25) -- (5,4.25) to[out=-90,in=90] (6,3.5) to[out=-90,in=90] (5,2.75) -- (5,1.75);
\draw [-,very thick,gray] (7,5.25) -- (7,4.25) to[out=-90,in=90] (6,3.5) to[out=-90,in=90] (7,2.75) -- (7,1.75);

\node[interaction] (h) at (6,3.5) {};
\node[Lax] (L5) at (5,2) {};
\node[Lax] (L6) at (7,2) {};
\node[Lax] (L7) at (5,5) {};
\node[Lax] (L8) at (7,5) {};
\path[-,very thick, color=black] (4,2) edge node[right] {} (L5)
			(L5) edge node[left] {} (L6);
\path[-,very thick, color=black] (L6) edge node[left] {} (8,2);
\path[-, very thick, color=black] (4,5) edge node[right] {} (L7)
			(L7) edge node[left] {} (L8);
\path[-,very thick, color=black] (L8) edge node[left] {} (8,5);

\path[-,thick, color=red] (8.25,3.5) edge node[left] {} (8.75,3.5);
\path[-,thick, color=red] (8.5,3.75) edge node[left] {} (8.5,3.25);

\path[-,very thick,gray] (10,0) edge node[left] {} (10,7);
\path[-,very thick,gray] (12,0) edge node[left] {} (12,7);

\node[dLax] (L9) at (10,2) {};
\node[Lax] (L10) at (12,2) {};
\node[Lax] (L11) at (10,5) {};
\node[Lax] (L12) at (12,5) {};
\path[-,very thick, color=black] (9,2) edge node[right] {} (L9)
			(L9) edge node[left] {} (L10);
\path[-,very thick, color=black] (L10) edge node[left] {} (13,2);
\path[-, very thick, color=black] (9,5) edge node[right] {} (L11)
			(L11) edge node[left] {} (L12);
\path[-,very thick, color=black] (L12) edge node[left] {} (13,5);

\path[-,thick, color=red] (13.25,3.5) edge node[left] {} (13.75,3.5);

\path[-,very thick,gray] (15,0) edge node[left] {} (15,7);
\path[-,very thick,gray] (17,0) edge node[left] {} (17,7);

\node[Lax] (L13) at (15,2) {};
\node[dLax] (L14) at (17,2) {};
\node[Lax] (L15) at (15,5) {};
\node[Lax] (L16) at (17,5) {};
\path[-,very thick, color=black] (14,2) edge node[right] {} (L13)
			(L13) edge node[left] {} (L14);
\path[-,very thick, color=black] (L14) edge node[left] {} (18,2);
\path[-, very thick, color=black] (14,5) edge node[right] {} (L15)
			(L15) edge node[left] {} (L16);
\path[-,very thick, color=black] (L16) edge node[left] {} (18,5);

\path[-,thick, color=red] (18.25,3.6) edge node[left] {} (18.75,3.6);
\path[-,thick, color=red] (18.25,3.4) edge node[left] {} (18.75,3.4);


\draw[-,very thick,gray] (20,7) to[out=-90,in=90] (21,6.25) to[out=-90,in=90] (20,5.5);
\draw[-,very thick,gray] (22,7) to[out=-90,in=90] (21,6.25) to[out=-90,in=90] (22,5.5);
\draw[-,very thick,gray] (20,0) -- (20,5.5);
\draw[-,very thick,gray] (22,0) -- (22,5.5);
\node[interaction] (h) at (21,6.25) {};
\node[Lax] (L17) at (20,2) {};
\node[Lax] (L18) at (22,2) {};
\node[Lax] (L19) at (20,5) {};
\node[Lax] (L20) at (22,5) {};
\path[-,very thick, color=black] (19,2) edge node[right] {} (L17)
			(L17) edge node[left] {} (L18);
\path[-,very thick, color=black] (L18) edge node[left] {} (23,2);
\path[-, very thick, color=black] (19,5) edge node[right] {} (L19)
			(L19) edge node[left] {} (L20);
\path[-,very thick, color=black] (L20) edge node[left] {} (23,5);

\path[-,thick, color=red] (23.25,3.5) edge node[left] {} (23.75,3.5);
\path[-,thick, color=red] (23.5,3.75) edge node[left] {} (23.5,3.25);

\path[-,very thick,gray] (25,0) edge node[left] {} (25,7);
\path[-,very thick,gray] (27,0) edge node[left] {} (27,7);

\node[dLax] (L21) at (25,2) {};
\node[Lax] (L22) at (27,2) {};
\node[Lax] (L23) at (25,5) {};
\node[Lax] (L24) at (27,5) {};
\path[-,very thick, color=black] (24,2) edge node[right] {} (L21)
			(L21) edge node[left] {} (L22);
\path[-,very thick, color=black] (L22) edge node[left] {} (28,2);
\path[-, very thick, color=black] (24,5) edge node[right] {} (L23)
			(L23) edge node[left] {} (L24);
\path[-,very thick, color=black] (L24) edge node[left] {} (28,5);

\path[-,thick, color=red] (28.25,3.5) edge node[left] {} (28.75,3.5);
\path[-,thick, color=red] (28.5,3.75) edge node[left] {} (28.5,3.25);

\path[-,very thick,gray] (30,0) edge node[left] {} (30,7);
\path[-,very thick,gray] (32,0) edge node[left] {} (32,7);

\node[Lax] (L25) at (30,2) {};
\node[Lax] (L26) at (32,2) {};
\node[dLax] (L27) at (30,5) {};
\node[Lax] (L28) at (32,5) {};
\path[-,very thick, color=black] (29,2) edge node[right] {} (L25)
			(L25) edge node[left] {} (L26);
\path[-,very thick, color=black] (L26) edge node[left] {} (33,2);
\path[-, very thick, color=black] (29,5) edge node[right] {} (L27)
			(L27) edge node[left] {} (L28);
\path[-,very thick, color=black] (L28) edge node[left] {} (33,5);

\path[-,thick, color=red] (33.25,3.5) edge node[left] {} (33.75,3.5);

\path[-,very thick,gray] (35,0) edge node[left] {} (35,7);
\path[-,very thick,gray] (37,0) edge node[left] {} (37,7);

\node[Lax] (L29) at (35,2) {};
\node[Lax] (L30) at (37,2) {};
\node[Lax] (L31) at (35,5) {};
\node[dLax] (L32) at (37,5) {};
\path[-,very thick, color=black] (34,2) edge node[right] {} (L29)
			(L29) edge node[left] {} (L30);
\path[-,very thick, color=black] (L30) edge node[left] {} (38,2);
\path[-, very thick, color=black] (34,5) edge node[right] {} (L31)
			(L31) edge node[left] {} (L32);
\path[-,very thick, color=black] (L32) edge node[left] {} (38,5);

\path[-,thick, color=red] (38.25,3.5) edge node[left] {} (38.75,3.5);

\path[-,very thick,gray] (40,0) edge node[left] {} (40,7);
\path[-,very thick,gray] (42,0) edge node[left] {} (42,7);

\node[Lax] (L33) at (40,2) {};
\node[dLax] (L34) at (42,2) {};
\node[Lax] (L35) at (40,5) {};
\node[Lax] (L36) at (42,5) {};
\path[-,very thick, color=black] (39,2) edge node[right] {} (L33)
			(L33) edge node[left] {} (L34);
\path[-,very thick, color=black] (L34) edge node[left] {} (43,2);
\path[-, very thick, color=black] (39,5) edge node[right] {} (L35)
			(L35) edge node[left] {} (L36);
\path[-,very thick, color=black] (L36) edge node[left] {} (43,5);

\end{tikzpicture}
\caption{A schematic depiction of $h^{n|m}(\LL_{\Lambda}(z) \stimes \LL_{\Lambda}(z))$, 
representing the local action of the Hamiltonian $H^{n|m}$ on the density operator $\rho_{\infty}=\Omega_{N}\Omega^{\dagger}_{N}$
(the coloring adopted from Figure \ref{fig:Sutherland}, and spectral and representation parameters are suppressed for clarity).
The process of brining the interaction $h^{n|m}$ across the horizontal legs generates terms which can be
interpreted as a operator divergence condition for two-row Lax operators $\mathbb{L}_{\Lambda}(z)$.}
\label{fig:two-leg}
\end{figure}
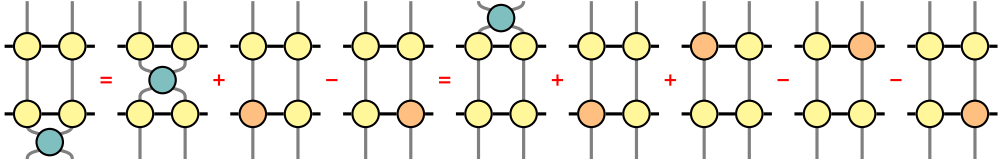

\subsection{Boundary compatibility condition}
\label{sec:boundary_conditions}

Given the Hamiltonian $H^{n|m}$ and a set of boundary dissipators $\mathcal{D}$, the fixed-point
condition \eqref{eqn:fixed_point} imposes a certain type of bulk-boundary matching condition.
It can be inferred from expression \eqref{eqn:divergence} that the fixed point condition $\mathcal{L}\,\rho_{\infty}=0$ admits a 
solution $\rho_{\infty}$ if and only if there exist an $\Omega$-amplitude (which amounts to find
the $\bL$-operator and the vacuum state $\rvac$) for which the dissipator $\mathcal{D}$ exactly cancels out
the right hand-side of Eq.~\eqref{eqn:divergence}. By plugging in a trial off-shell density operator $\rho_{\Lambda}(z)$
and demanding the \emph{on-shell} condition one obtains a system of boundary algebraic equations for the undetermined representation 
parameters which depends also on the physical coupling parameters $\g$ of the reservoirs. The solution, when it exists,
singles out a unique density operator $\rho_{\infty}(\g)$.

Combining Eq.~\eqref{eqn:ss_universal} with a general solution of the bulk condition \eqref{eqn:Sutherland_general}
results in two decoupled sets of \emph{boundary compatibility conditions}, which can be cast in the compact form~\cite{IP14}
\begin{equation}
\begin{split}
\llvac \Big(\mathcal{D}_{\rm L}+\ii \partial_{z}\Big)\LL_{\Lambda}(z) &= 0,\\
\Big(\mathcal{D}_{\rm R}-\ii \partial_{z}\Big) \LL_{\Lambda}(z) \rrvac &= 0.
\label{eqn:boundary_conditions}
\end{split}
\end{equation}
The boundary conditions of this form generically yields an \emph{overdetermined} system of equations for the free
parameters of the two-row Lax operator $\LL_{\Lambda}(z)$.
Indeed, it is not difficult to confirm that in spite of integrability of the bulk interactions \emph{generic} boundary dissipators
do not lead to any solutions of Eqs.~\eqref{eqn:boundary_conditions}.
In other words, for some general choice of boundary dissipators there exist no off-shell operator $\rho_{\Lambda}(z)$
which would satisfy the fixed-point condition of Eq.~\eqref{eqn:fixed_point}. Of course this should not be surprising at all since
typical dissipation processes result in a `non-integrable' Liouvillian dynamics in which a na\"{i}ve separation of bulk and boundary 
parts cannot be justified. Needless to say that in such a case there exists no obvious explicit representation of the steady states 
either. It is therefore quite remarkable that integrable lattice models with $\mathfrak{su}(n|m)$-symmetric interactions $h^{n|m}$ do 
allow for certain elementary (so-called integrable) boundary dissipators which lead to non-trivial solutions to boundary
equations \eqref{eqn:boundary_conditions}.

\begin{figure}[ht]
\centering
\begin{tikzpicture}[scale=0.5]
\tikzstyle{Lax} = [circle, thick, color=black, minimum width=18pt, fill=yellow!50, draw, inner sep=0pt]
\tikzstyle{dLax} = [circle, thick, color=black, minimum width=18pt, fill=orange!50, draw, inner sep=0pt]
\tikzstyle{jump} = [circle, thick, color=black, minimum width=18pt, fill=purple!50, draw, inner sep=0pt]

\draw[->|,very thick,black] (-1.25,0) -- (1.25,0);
\draw[->|,very thick,black] (-1.25,2) -- (1.25,2);
\draw[->,very thick,gray] (0,-3) -- (0,5.5);
\node[jump] (A) at (0,-2) {\scriptsize $2A$};
\node[Lax] (L1) at (0,0) {\scriptsize ${\bf L}$};
\node[Lax] (L2) at (0,2) {\scriptsize ${\bf \overline{L}}$};
\node[jump] (Ad) at (0,4) {\tiny $A^{\dagger}$};

\path[-,thick, color=red] (1.75,1) edge node[left] {} (2.25,1);

\draw[->|,very thick,black] (2.75,0) -- (5.25,0);
\draw[->|,very thick,black] (2.75,2) -- (5.25,2);
\draw[->,very thick,gray] (4,-3) -- (4,4);
\draw[-,very thick,gray] (4,3.9) -- (4,5.5);
\node[jump] (AdA1) at (4,-2) {\tiny $A^{\dagger}A$};
\node[Lax] (L3) at (4,0) {\scriptsize ${\bf L}$};
\node[Lax] (L4) at (4,2) {\scriptsize ${\bf \overline{L}}$};

\path[-,thick, color=red] (5.75,1) edge node[left] {} (6.25,1);

\draw[->|,very thick,black] (6.75,0) -- (9.25,0);
\draw[->|,very thick,black] (6.75,2) -- (9.25,2);
\draw[->,very thick,gray] (8,-3) -- (8,-2);
\draw[-,very thick,gray] (8,-2.1) -- (8,5.5);
\node[Lax] (L5) at (8,0) {\scriptsize ${\bf L}$};
\node[Lax] (L6) at (8,2) {\scriptsize ${\bf \overline{L}}$};
\node[jump] (AdA2) at (8,4) {\tiny $A^{\dagger}A$};

\path[-,thick, color=red] (9.75,0.9) edge node[left] {} (10.25,0.9);
\path[-,thick, color=red] (9.75,1.1) edge node[left] {} (10.25,1.1);

\draw[->|,very thick,black] (10.75,0) -- (13.25,0);
\draw[->|,very thick,black] (10.75,2) -- (13.25,2);
\draw[->,very thick,gray] (12,-3) -- (12,-2);
\draw[->,very thick,gray] (12,-2.1) -- (12,4);
\draw[-,very thick,gray] (12,3.9) -- (12,5.5);
\node[dLax] (L7) at (12,0) {\scriptsize $\ii{\bf L}^{\prime}$};
\node[Lax] (L8) at (12,2) {\scriptsize ${\bf \overline{L}}$};

\path[-,thick, color=red] (13.75,1) edge node[left] {} (14.25,1);
\draw[-,thick, color=red] (14,0.75) edge node[left] {} (14,1.25);

\draw[->|,very thick,black] (14.75,0) -- (17.25,0);
\draw[->|,very thick,black] (14.75,2) -- (17.25,2);
\draw[->,very thick,gray] (16,-3) -- (16,-2);
\draw[->,very thick,gray] (16,-2.1) -- (16,4);
\draw[-,very thick,gray] (16,4) -- (16,5.5);
\node[Lax] (L7) at (16,0) {\scriptsize ${\bf L}$};
\node[dLax] (L8) at (16,2) {\scriptsize $\ii {\bf \overline{L}}^{\prime}$};

\end{tikzpicture}
\caption{Graphical interpretation of the boundary compatibility condition as given by equation \eqref{eqn:boundary_conditions}
displayed for the right boundary at lattice site $N$. The left-hand side shows schematically
the action of the dissipator $\mathcal{D}$ on the $\mathbb{L}$-operator decomposed into three terms which
the define the action of the jump operator $A_{N}$. The termination point of the horizontal arrow signifies the contraction with
the right auxiliary vacuum state. Note that the boundary condition has to be satisfied for all values of physical indices.}
\end{figure}
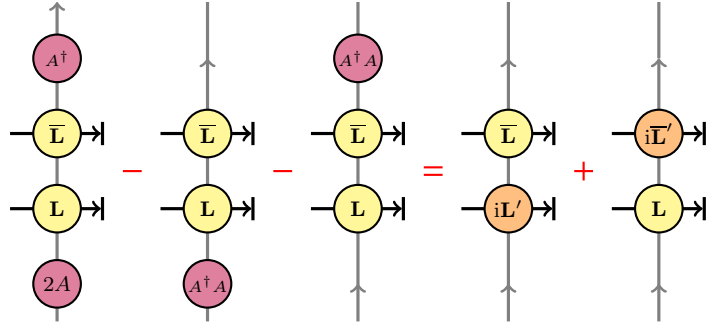

\subsection{Integrable dissipative boundaries}
\label{sec:integrable_boundaries}

We consider a pair of dissipative boundary processes which involves any (but arbitrary) pair
of states from the local Hilbert space $\mathbb{C}^{n|m}$. Denoting them by $\ket{\alpha}$ and 
$\ket{\beta}$, we posit the jump operators of the form\footnote{In principle the left 
and right reservoirs can be assigned unequal couplings without spoiling integrability (see e.g. \cite{Prosen_review}).
In this work we prefer for simplicity to concentrate to the situation with equal coupling rates.}
\begin{equation}
A_{1} = \sqrt{g}\,E^{\alpha \beta}_{1},\qquad A_{N} = \sqrt{g}\,E^{\beta \alpha}_{N},
\label{eqn:integrable_boundaries}
\end{equation}
parametrized by a single reservoir coupling parameter $g$. Since Lindblad dissipators which enter in
Eq.~\eqref{eqn:integrable_boundaries} operate non-trivially only on the boundary sites  
of the chain, the jump operator from Eq.~\eqref{eqn:integrable_boundaries} can be interpreted as a source and drain associated
to $U(1)$ particle currents.

In models with multiple states per site such as $\mathfrak{su}(n|m)$ chains considered here, diagonal
`density operators' $E^{aa}_{i}$ obey the following local continuity equations
\begin{equation}
\partial_{t}\big(E^{aa}_{i}-E^{bb}_{i}\big) = \ii \big[H,E^{aa}_{i}-E^{bb}_{i}\big] = j^{ab}_{i-1,i} - j^{ab}_{i,i+1},
\end{equation}
where $j^{ab}_{i}$ denote partial currents between two level $\ket{a}$ and $\ket{b}$ locally at lattice site $i$.
Total current densities between two adjacent lattice sites are then obtained by summing over all partial currents, that is 
$j^{a}_{i,i+1}\equiv \sum_{b=1}^{n}j^{ab}_{i,i+1}$, and fulfil
\begin{equation}
\partial_{t}E^{aa}_{i} = \ii \big[H,E^{aa}_{i}\big]= j^{a}_{i-1,i} - j^{a}_{i,i+1}.
\end{equation}
Integrable $\mathfrak{su}(n|m)$ symmetric Hamiltonians $H^{n|m}$ conserve each total particle number
$N^{a}=\sum_{i}E^{aa}_{i}$ independently, i.e. $[H^{n|m},N^{a}]=0$. The addition of dissipation however destroys
the conservation of $N^{a}$ if $a\in \{\alpha,\beta\}$.

To better examine this situation, we notice that a particular choice of dissipative boundary condition as 
given by Eq.~\eqref{eqn:integrable_boundaries} allows to decompose Liouvillian dynamics into invariant subspaces,
\begin{equation}
\mathcal{H} = \bigoplus_{\underline{\nu}} \mathcal{H}^{\underline{\nu}},\qquad
\underline{\nu}\equiv (\nu_{1},\ldots,\nu_{n+m}),
\label{eqn:foliation}
\end{equation}
where orthogonal Hilbert subspaces $\mathcal{H}^{\underline{\nu}}$ are defined via
$N^{\gamma}\mathcal{H}^{\underline{\nu}} = \nu_{\gamma} \mathcal{H}^{\underline{\nu}}$, with eigenvalues
$\nu_{\gamma}\in \{0,1,\ldots,N\}$ for $\gamma \in \{1,\ldots,n+m\}$. Accordingly, we introduce endomorphisms
$\mathcal{O}^{\underline{\nu}}={\rm End}(\mathcal{H}^{\underline{\nu}})$, i.e. linear spaces of operators operating on
$\mathcal{H}^{\underline{\nu}}$. This means that states from $\mathcal{H}^{\underline{\nu}}$ have
well-defined values of all particle number operators $N^{\gamma}$.
When ${\rm rank}(\mathfrak{g})>1$, there exist \emph{at least one} number operator $N^{\gamma}$ such that
\begin{equation}
\big[H^{n|m},N^{\gamma}\big] = \big[A_{1},N^{\gamma}\big] = \big[A_{N},N^{\gamma}\big]=0.
\label{eqn:strong_symmetry}
\end{equation}
This is an example of the so-called `\emph{strong Liouvillian symmetry}'~\cite{Beri}. In fact, all $N^{\gamma}$ correspond to
strong symmetries, with the exception of the two distinguished indices which belong to a pair of levels affected by the boundary 
dissipation, that is $\gamma\in \{\alpha,\beta\}$. This immediately implies degeneracy\footnote{Uniqueness of individual steady state 
components from each conserved subspace follows from the theorem of Evans~\cite{Evans77}.} of the
steady states (cf.~\cite{IP14}) and vanishing current expectation values $\langle j^{\gamma} \rangle_{\infty}=0$.
Thus only current densities $\langle j^{\alpha} \rangle_{\infty}$ and  $\langle j^{\beta} \rangle_{\infty}$ 
can take non-vanishing steady-state expectation values.

When dealing with degenerate null spaces of the generator $\mathcal{L}$,
the steady state operator $\rho_{\infty}$ naturally decomposes in terms of independent fixed-point components
$\rho^{\underline{\mu}}_{\infty}$ from individual invariant subspaces $\mathcal{O}^{\underline{\mu}}$, with
$\underline{\mu}=\underline{\nu} \setminus \{\nu_{\alpha},\nu_{\beta}\}$.
That is, we have
\begin{equation}
\rho_{\infty} = \sum_{\underline{\mu}}\rho^{\underline{\mu}}_{\infty},\qquad
\rho^{\underline{\mu}}_{\infty} = \mathcal{P}^{\underline{\mu}}\,\rho_{\infty},
\end{equation}
with $\mathcal{P}^{\underline{\mu}}$ denoting orthogonal projectors onto subspaces $\mathcal{O}^{\underline{\mu}}$.
Because each invariant component $\rho^{\underline{\mu}}_{\infty}$ satisfies $\mathcal{L} \rho^{\underline{\mu}}_{\infty}=0$ for
all allowed values of particle numbers $\underline{\mu}$, they may be combined in any convex-linear combination
\begin{equation}
\rho_{\infty} = \sum_{\underline{\mu}}c_{\underline{\mu}}\,\rho^{\underline{\mu}}_{\infty},
\label{eqn:grand_canonical}
\end{equation}
with $c_{\underline{\mu}}$ representing a $(n+m-2)$-component vector with non-negative components.
The steady-state operator $\rho_{\infty}$ as defined by Eq.~\eqref{eqn:grand_canonical} can be thus regarded as a grand canonical 
nonequilibrium ensemble with coefficients $c_{\underline{\mu}}$, which in analogy to grand canonical ensembles have the role of 
particle chemical potentials.
Notice that Eq.~\eqref{eqn:grand_canonical} can be conveniently cast in the form of a matrix-product operator along the
lines of ref. \cite{IP14} for the $\mathfrak{su}(3)$ chain.

A heuristic analysing of `integrable boundaries' specified Eq.~\eqref{eqn:integrable_boundaries}, e.g. by computing exact
solutions for quantum chain of small length, reveals that the fixed point is a non-trivial current-carrying steady states of
particularly simple structure. In the remainder of the paper we demonstrate that the steady-state solutions exhibit a particular 
algebraic representation which directly link to fundamental objects of quantum algebras.
It is worthwhile stressing nonetheless that, despite the simplicity of our effective reservoirs, the entire spectrum of 
$\mathcal{L}$ -- typically referred to as the \emph{Liouville decay modes} -- remains highly complex and lack any obvious structure. 
With that said, it is therefore only the fixed-point solutions $\rho_{\infty}$ of Eq.~\eqref{eqn:fixed_point} which admit an exact 
description. It is also instructive to remark here that even the integrable steady state density operators themselves do not enjoy
the full quantum group symmetry of the underlying theory.
Indeed, as a consequence of the foliation \eqref{eqn:foliation} of the Lindbladian flow, the global 
residual symmetry of $\rho_{\infty}$ is merely $U(1)^{\otimes n+m-2}$. However, as subsequently demonstrated,
the \emph{local} symmetry of the $\Omega$-amplitude is much larger.
The symmetry content of the steady state solution will be carefully examined in the next sections. Particularly,
the local symmetry of the $\Omega$-amplitudes will become apparent on the basis of previously discussed Lax representation
(see also Section~\ref{sec:vacuumQ} for additional remarks).

\section{Graded Yangians}
\label{sec:Yangians}

Yangians are certain infinite-dimensional quadratic associative algebras which belong to a class of (quasi-triangular) Hopf algebras, 
widely referred to as quantum groups. Yangians can be defined in various equivalent ways~\cite{Jimbo85,Drinfeld86,Drinfeld88}.
Here we employ the `FRT realization'~\cite{FRT} (also known as the `RTT realization'), in which Yang--Baxter
equation \eqref{eqn:graded_YB} takes a role of the defining relation.

We specialize the discussion to Yangians $\mathcal{Y}\equiv Y(\mathfrak{g})$ of Lie superalgebras
$\mathfrak{g}=\mathfrak{gl}(n|m)$~\cite{Nazarov91,Molev}. Recall that the signature $n|m$ indicates
that the local Hilbert space consists of $n$ bosonic and $m$ fermionic states.
Generators of $\mathcal{Y}$ are given as the operator-valued coefficients of the Lax operator $\bL(z)$ expanded as
a formal Laurent series
\begin{equation}
\bL^{ab}(z) = \bL^{ab}_{(0)}+z^{-1}\,\bL^{ab}_{(1)}+z^{-2}\,\bL^{ab}_{(2)}+\ldots.
\end{equation}
By imposing Yang--Baxter equation \eqref{eqn:graded_YB} as the defining relation,
we obtain an infinite set of quadratic algebraic conditions
\begin{equation}
\Big[\bL^{ab}_{(r)},\bL^{cd}_{(s)}\Big] =
(-1)^{ab+ac+bc}\sum_{i=1}^{{\rm min}(r,s)}
\Big(\bL^{cb}_{(r+s-i)}\bL^{ad}_{(i-1)} - \bL^{cb}_{(i-1)}\bL^{ad}_{(r+s-i)}\Big).
\label{eqn:Yangian_defining}
\end{equation}
The level-$0$ generators $\bL^{ab}_{(0)}$ are scalars belonging to the center of $\mathcal{Y}$.
In the scope of our application, we are only be interested in the class of fundamental \emph{rational} solutions
of Eq.~\eqref{eqn:graded_R} which are of degree one in the spectral parameter $z$,\footnote{Realizations of $\mathcal{Y}$ which are of 
higher degree in $z$ have been briefly discussed in \cite{Frassek11}. At the moment it remains unclear to us whether these solutions 
are important in the studied setup.}
\begin{equation}
\bL^{ab}(z) = \bL^{ab}_{(0)} + z^{-1}\,\bL^{ab}_{(1)},\qquad \bL^{ab}_{(k)}\equiv 0\quad {\rm for}\quad k\geq 2.
\end{equation}
This choice represents, in mathematical terms, an evaluation homomorphism from the Yangian to the universal enveloping algebra of
$\mathfrak{g}$, $\mathcal{Y}(\mathfrak{g})\mapsto \mathcal{U}(\mathfrak{g})$. With this restriction, representations of
$\mathcal{Y}$ are in one-to-one correspondence with representations of the classical Lie (super)algebra $\mathfrak{g}$.

\paragraph{Automorphisms.}
It is instructive to shorty discuss the gauge freedom due to automorphisms of $\mathcal{Y}$,
i.e. transformations which preserve the algebra \eqref{eqn:Yangian_defining} (cf. \cite{Bazhanov11}).
These comprise of (i) rescaling  $\bL(z)$ with an arbitrary complex-valued scalar 
function $f(z)$, (ii) shifting the spectral parameter $z\to z+z^{\prime}$ and
(iii) applying a $(n+m)$-dimensional $GL(n|m)$ gauge transformations which acts in $\mathbb{C}^{n|m}$ and
is given by two arbitrary invertible matrices $G_{L}$ and $G_{R}$, $\bL(z)\to G_{L}\,\bL(z)\,G_{R}$.
In addition, there exist anti-automorphisms of $\mathcal{Y}$, i.e. transformations which only preserve the defining
relations \eqref{eqn:graded_YB} up to exchanging the order of tensor factors. Examples of these are transposition of the matrix 
space $\bL(z)\to \bL^{t}(z)$, and reflection in the spectral plane $\bL(z)\to \bL(-z)$.
Any composition of two anti-automorphisms is again an automorphism.

\paragraph{Rank--degenerate realizations.}
The list of transformations given above nonetheless do not exhaust all possibilities of realizing $\mathcal{Y}$. As pointed out
in \cite{Bazhanov11,Frassek11}, equation \eqref{eqn:graded_YB} admits a class of `degenerate solutions' provided one
relaxes the requirement $\bL_{(0)}=1$. This is a viable choice because the level-$0$ generators 
$\bL^{ab}_{(0)}$ are central and can therefore take arbitrary (possibly vanishing) values. We may thus quite generally prescribe
\begin{equation}
\bL_{(0)} = {\rm diag}(\underbrace{1,1,\ldots,1}_{|I|},\underbrace{0,\ldots,0}_{|\overline{I}|}),\qquad 
1\leq |I| \leq n+m,
\label{eqn:signature}
\end{equation}
modulo equivalent choices which correspond to permutations of $0$s and $1$s.
Such a restriction obviously induces another block structure\footnote{We follow the notation of \cite{Bazhanov11,Frassek11} and employ 
the two-index labelling of the Yangian generators.} on $\mathbb{C}^{n|m}$ under which the generators of $\mathcal{Y}$ split as
\begin{equation}
\bL_{(1)} =
\begin{pmatrix}
\bA^{ab} & \bB^{a \dot{b}} \\
\bC^{\dot{a}b} & \bD^{\dot{a}\dot{b}}
\end{pmatrix}.
\label{eqn:block_structure}
\end{equation}
The ranges of ordinary (undotted) and dotted indices are
\begin{equation}
a,b\in I=\{1,2,\ldots,|I|=p+q\},\qquad \dot{a},\dot{b}\in \overline{I}=\{1,\ldots,n+m\}\backslash I,
\end{equation}
where $p$ ($q$) denotes the number of bosonic (fermionic) states in the index set $I$. Similarly, we shall denote by
$\dot{p}$ ($\dot{q}$) is the number of bosonic (fermionic) states contained in the complementary set $\overline{I}$.
The defining relations of the resulting `hybrid algebra' $\mathfrak{A}^{I}_{n,m}$ are readily obtained
by plugging Eq.~\eqref{eqn:block_structure} in Eq.~\eqref{eqn:Yangian_defining}, and select the level-$0$ generators in accordance
with Eq.~\eqref{eqn:signature}.
in accordance with the prescription of Eq.~\eqref{eqn:signature}, i.e. $\bL^{ab}_{(0)}=\delta_{aI}\,\delta_{bI}$.
Since the generators $\bD^{\dot{a}\dot{b}}$ are central, it is convenient to pick a gauge by
setting $\bD^{\dot{a}\dot{b}}=\delta_{\dot{a}\dot{b}}$. The remaining non-trivial commutation relations read
\begin{align}
\label{eqn:hybrid_algebra}
\Big[\bA^{ab},\bA^{cd}\Big] &= (-1)^{ab+ac+bc}(\delta_{ad}\,\bA^{cb}- \delta_{cb}\,\bA^{ad}),&\quad
\Big[\bA^{ab},\bB^{c\dot{d}}\Big] &= -(-1)^{ab+ac+bc}\delta_{cb}\,\bB^{a\dot{d}}, \nonumber \\
\Big[\bA^{ab},\bC^{\dot{c}d}\Big] &= (-1)^{ab+a\dot{c}+b\dot{c}}\delta_{ad}\,\bC^{\dot{c}b},&\quad
\Big[\bB^{a\dot{b}},\bC^{\dot{a}b}\Big] &= (-1)^{\dot{a}}\delta_{ab}\,\delta_{\dot{a}\dot{b}},\\
\Big[\bB^{a\dot{b}},\bB^{c\dot{s}}\Big] &= 0,&\quad \scom{\bC^{\dot{a}b}}{\bC^{\dot{c}s}} &= 0.\nonumber
\end{align}

\subsection{Oscillator realizations}

Commutation relations \eqref{eqn:hybrid_algebra} have been derived in \cite{Bazhanov11,Frassek11}, where the authors
provide a realization in terms of $\mathfrak{gl}(p|q)$ `super spin' generators $\bJ^{ab}$ (for $a,b\in I$, and with $p+q=|I|$),
\begin{equation}
\Big[\bJ^{ab},\bJ^{cd}\Big] = \delta_{cb}\,\bJ^{ad} - (-1)^{(a+b)(c+d)}\delta_{ad}\,\bJ^{cb},
\end{equation}
and additional $|I|\cdot |\overline{I}|$ canonical bosonic or fermionic oscillators which obey graded canonical
commutation relations
\begin{equation}
\Big[\bxi^{\dot{a}b},\bxid^{c\dot{d}}\Big] = \delta_{cb}\,\delta_{\dot{a}\dot{d}}.
\end{equation}
where a generator $\bxid^{a\dot{b}}$ should be understood as a creation operator of a bosonic (fermionic) oscillator if $p(\dot{b})=0$ 
($p(\dot{b})=1$), for $a\in I$ and $\dot{b}\in \overline{I}$. The oscillator part of the algebra $\mathfrak{A}^{I}_{n,m}$,
denoted by $\mathfrak{osc}(p+q|\overline{p}+\overline{q})$, is associated with a multi-component Fock space
$\mathcal{B}^{\otimes (p+\dot{p})}\otimes \mathcal{F}^{\otimes (q+\dot{q})}$, where each factor $\mathcal{B}$ ($\mathcal{F}$) belongs 
to an irreducible bosonic (fermionic) Fock space.
In terms of these `super spins' and `super oscillators', the level-$1$ generators $\bL^{ab}_{(1)}$ take the \emph{canonical} form
\begin{equation}
\begin{split}
\bA^{ab} &= -(-1)^{b}\big(\bJ^{ab} + \mathbf{N}^{ab}\big),\\
\bB^{a\dot{b}} &= \bxid^{a\dot{b}},\\
\bC^{\dot{a}b} &= -(-1)^{b}\bxi^{\dot{a}b},\\
\bD^{\dot{a}\dot{b}} &= \delta_{\dot{a}\dot{b}},
\end{split}
\label{eqn:canonical_form}
\end{equation}
where
\begin{equation}
\mathbf{N}^{ab} = \sum_{\dot{d}\in \overline{I}}\bxid^{a\dot{d}}\bxi^{\dot{d}b}+\tfrac{1}{2}(-1)^{a+\dot{d}}\delta_{ab}.
\end{equation}

\subsection{Partonic Lax operators}
For a Lie superalgebra $\mathfrak{g}$ of ${\rm rank}(\mathfrak{g})=n+m$, there are in
total $2^{n+m}$ distinct types of hybrid-type subalgebras $\mathfrak{A}^{I}_{n,m}$. The latter
are in a bijective correspondence with all possible choices of set $I$. Notice that
this counting we are excluding the various possibilities of choosing the grading.
By additionally excluding permutation equivalent choices, we eventually deal with
finite-dimensional Lie superalgebras of the type $\mathfrak{gl}(p|q)\otimes \mathfrak{osc}(p+\dot{p}|q+\dot{q})$.

The simplest rank-degenerate solutions of the graded Yang--Baxter algebra~\eqref{eqn:graded_YB}
belongs to the single-indexed sets $|I|=1$ and were dubbed in \cite{Bazhanov10} as the \emph{partonic} solutions.
These consist solely from $n+m-1$ oscillators arranged in a distinctive cross-shaped form,
\begin{equation}
\bL_{\{a\}}(z) =
\begin{pmatrix}
1 & & & -(-1)^{a}\bxi^{1,a} & & & \cr
& \ddots & & \vdots & & & \cr
& & 1 & -(-1)^{a}\bxi^{a-1,a} & & & \cr
\bxid^{a,1} & \ldots & \bxid^{a,a-1} & z - \mathbf{N}^{a}_{\overline{I}} &
\bxid^{a,a+1} & \ldots & \bxid^{a,n+m} \cr
 & & & -(-1)^{a}\bxi^{a+1,a} & 1 & & \cr
 & & & \vdots & & \ddots & \cr
 & & & -(-1)^{a}\bxi^{n+m,a} & & & 1
\end{pmatrix},
\label{eqn:partonic}
\end{equation}
with
\begin{equation}
\mathbf{N}^{a}_{\overline{I}}=\sum_{\dot{b}\in \overline{I}}(-1)^{\dot{b}}\left(\bxid^{a\dot{b}}\bxi^{\dot{b}a}+
\tfrac{1}{2}(-1)^{a+\dot{b}}\right).
\end{equation}
Here the integers $a\in \{1,\ldots,n+m\}$ in the subscript of $\bL_{\{a\}}(z)$ are being used to indicate the location of
the single non-vanishing level-$0$ generator, cf. Eq.~\eqref{eqn:signature}. As shown in appendix \ref{app:fusion},
all Lax operators associated to $\mathfrak{A}^{|I|\geq 2}_{m,n}$ can be systematically generated from 
the partonic solutions which carry $\mathfrak{A}^{|I|=1}_{n,m}$ by employing a universal fusion formula, yielding
`multi-partonic' Lax operators which are equivalent to canonical Lax operators given by Eq.~\eqref{eqn:canonical_form}.

\section{Exact steady states for integrable quantum spin chains}
\label{sec:solutions}

In this section we finally present a few explicit examples for the steady-state solutions of the boundary-driven $\mathfrak{su}(n|m)$
quantum chains, subjected to integrable dissipative boundaries given by \eqref{eqn:integrable_boundaries}. As the first step, we
account for kinematic constraints and construct off-shell density operators which take the universal form of
equation \eqref{eqn:ss_universal}. Subsequently, the goal is to find an appropriate internal structure of the
Lax operator $\LL_{\boldsymbol{\Lambda}}$ and the auxiliary vacuum state $\rvac$ which fulfil the requirements of the boundary 
equations \eqref{eqn:boundary_conditions}.

Notice first that there exist in total $2\times \binom{n+m}{2}$ ways of assigning the dissipators of 
Eq.~\eqref{eqn:integrable_boundaries}, representing all the possibilities of selecting a pair of target levels
$\ket{\alpha}$ and $\ket{\beta}$. The extra factor of $2$ reflects a possibility of exchanging $\ket{\alpha}$ with $\ket{\beta}$
which results in a state of opposite chirality, i.e. reverses directions of particle currents.
It turns out that every choice of $\ket{\alpha}$ and $\ket{\beta}$ leads to a solution to
Eq.~\eqref{eqn:fixed_point} which is uniquely characterized by specifying a representation and the corresponding labels for
the auxiliary algebra $\mathfrak{A}^{I}_{n,m}$ of the Lax operator $\LL_{\Lambda}(z)$.

\subsection{Fundamental integrable spin models}

The significance of partonic Lax operators and the structure of the steady-state solutions
is perhaps best illustrated by explicitly working out a few simplest examples.
To this end, we first consider the non-graded interactions, and initially examine the most studied case
of the $\mathfrak{su}(2)$ spin chain (the isotropic Heisenberg spin-$1/2$ model), with interaction density
$h^{2|0}=\sum_{a,b=1}^{2}E^{ab}\otimes E^{ba}$. Let us remark that this particular instance has been considered initially
in the seminal paper \cite{ProsenPRL107} where the solutions was found with a somewhat different approach, and afterwards
re-obtained in a more compact and symmetric form in \cite{KPS13}. The derivation from Yang--Baxter algebra has been presented
in \cite{IZ14}. Nevertheless, to uncover the connection with partonic Lax operators and embed this solution in a unified theoretic 
framework, we shall reproduce it below once again.

In the $\mathfrak{su}(2)$ spin chain, the local building block of the $\Omega$-amplitude is given by a
two-parametric Lax operator $\bL^{-}_{j}(z)$ acting on $\mathcal{V}_{\square}\otimes \mathcal{V}^{-}_{j}$,
whose auxiliary space $\mathcal{V}^{-}_{j}$ represents a \emph{lowest-weight} $\mathfrak{sl}(2)$ module
spanned by an infinite tower of states $\{\ket{k}\}_{k=0}^{\infty}$.
We adopt the $\mathfrak{sl}(2)$ spin generators obeying algebraic relations
\begin{equation}
\big[\bJ^{3},\bJ^{\pm}\big] = \pm\,\bJ^{\pm},\qquad \big[\bJ^{+},\bJ^{-}\big] = 2\,\bJ^{3},
\label{eqn:sl2_commutation}
\end{equation}
whose action on $\mathcal{V}^{-}_{j}$ is prescribed by
\begin{equation}
\bJ^{3}\ket{k} = (k-j)\ket{k},\quad
\bJ^{+}\ket{k} = (2j-k)\ket{k+1},\quad
\bJ^{-}\ket{k} = k\ket{k-1}.
\label{eqn:sl2_hw_module}
\end{equation}
State $\ket{0}$ has the lowest weight, $\bJ^{-}\ket{0}=0$, and will be referred to as the vacuum.

By recalling that all the solutions factorize in accordance with Eq.~\eqref{eqn:amplitude_factrorization}, the
off-shell Lax operator $\LL_{\Lambda}(z)$ which defines the steady-state solution $\rho_{\infty}$
is a product of two copies of auxiliary representations, cf. Eq.~\eqref{eqn:double_L}.
This factorization makes is possible to express the final solution by only specifying a pair of complex
parameters: a $\mathfrak{sl}(2)$ weight which is interpreted as a complex spin $j$, and a complex-valued spectral
parameter $z$. Specifically the two-parameteric off-shell Lax operator which we denote by $\mathbb{L}_{j}(z)$ is represented
in the following compact form
\begin{equation}
\LL_{j}(z) = \bL^{[1]-}_{j}(z)\,\bL^{[2]+}_{-j}(z).
\label{eqn:LL_sl2}
\end{equation}
The notation used is as follows. Integer indices in superscript brackets are used to assign an operator $\bL_{j}(z)$ into the
corresponding tensor factor in the multi-component auxiliary space. In addition, in the superscripts we also employed
extra parity signatures which are required to correctly specify the type of the $\mathfrak{sl}(2)$ module. Namely,
while the first auxiliary copy is realized in the lowest-weight module as prescribed
by Eq.~\eqref{eqn:sl2_hw_module}, the second factor in Eq.~\eqref{eqn:LL_sl2} must be associated with the highest-weight realization 
of $\mathfrak{sl}(2)$ algebra $\mathcal{V}^{+}_{j}$.
The highest-weight Lax operator $\bL^{+}_{j}(z)$ can be readily obtained from $\bL^{-}_{j}(z)$ by applying the
spin-reversal transformation
\begin{equation}
\mathcal{V}^{-}_{j}\to \mathcal{V}^{+}_{j}:\qquad \bJ^{\pm}\to \bJ^{\mp},\qquad \bJ^{3}\to -\bJ^{3}.
\label{eqn:spin-reversal}
\end{equation}
The highest weight state $\ket{0}$ from $\mathcal{V}^{+}_{j}$ is distinguished by $\bJ^{+}\ket{0}=0$.

Plugging the off-shell form of Eq.~\eqref{eqn:LL_sl2} into the boundary conditions \eqref{eqn:boundary_conditions} yields a system of 
polynomial equations with a unique solution
\begin{equation}
z= 0,\qquad j = \frac{\ii}{g}.
\end{equation}
Notice that the auxiliary vacuum state takes the product form, $\rrvac = \ket{0}\otimes \ket{0}$, and is determined by the internal 
structure of $\LL_{j}(z)$. In order to reverse the direction of driving we may simply exchange the target states
$\ket{\alpha}\leftrightarrow \ket{\beta}$. This amounts to interchange the factors in Eq.~\eqref{eqn:LL_sl2}.

Before proceeding with other examples, let us stress again that a proper identification of the internal structure of the
two-leg Lax operator $\LL_{j}(z)$ is crucial. Once the convention for labelling the irreducible $\mathfrak{sl}(2)$ Verma modules
is being fixed, there exist only one correct assignment of irreducible spaces (incorrect assignments produce a system of
boundary equations which admits no solution).

\paragraph{Asymmetric driving.}
Let us mention a simple trick which enables to generalize the solutions to the case of \emph{unequal} reservoir coupling constants.
Considering as an example the $\mathfrak{su}(2)$ spin chain, we may impose an asymmetric pair of Lindblad jump operators of 
the form
\begin{equation}
A_{1} = \sqrt{g/ \zeta}\,E^{12}_{1},\qquad
A_{N}=\sqrt{g\,\zeta}\,E^{21}_{N}.
\end{equation}
This choice yields an extended class of solutions which is connected to the special case of equal couplings
by a diagonal tilting transformation -- a one-parameter automorphism of $\mathcal{Y}$ -- by applying
\begin{equation}
\bL^{-}_{j}(z) \to \bL^{-}_{j}(z)
\begin{pmatrix}
\zeta & 0 \\
0 & 1
\end{pmatrix},
\end{equation}
on every local spin space $\mathbb{C}^{2}$.
The solution to the boundary compatibility conditions is then given by
\begin{equation}
z = \frac{\ii}{2}\left(\frac{1}{g\,\zeta}-\frac{\zeta}{g}\right),\qquad
j = \frac{\ii}{2}\left(\frac{1}{g\,\zeta}+\frac{\zeta}{g}\right).
\end{equation}

\paragraph{Models of higher-rank symmetry.}
The simplest higher-rank model is the $\mathfrak{su}(3)$ spin chain (with interaction $h^{3}$),
often called in the literature as the Lai--Sutherland model~\cite{Lai74,Sutherland75}.
Solutions to the fixed-point condition \eqref{eqn:fixed_point} have been originally identified and parametrized in \cite{IP14} and
now represent a \emph{degenerate} manifold of steady states. As discussed earlier in Section \ref{sec:integrable_boundaries},
degeneracy of the null space of $\mathcal{L}$ is a consequence of the conservation of the number operator $N^{\gamma}$ associated to a 
distinguished noise-protected state $\ket{\gamma}$ (which depends on $\overline{I}$).

The Lax operator for the $\Omega$-amplitude now operates on a space of three auxiliary particles,
a \emph{non-compact} $\mathfrak{sl}(2)$ spin and two species of canonical bosons.
Bosonic particles obey canonical oscillator algebra,
\begin{equation}
\big[\bb,\bb^{\dagger}\big]=1,\quad
\big[\bh,\bb^{\dagger}\big]=\bb^{\dagger},\quad
\big[\bh,\bb\big]=-\bb,\quad \bh \equiv \bb^{\dagger}\bb+\tfrac{1}{2},
\end{equation}
and live in the canonical Fock space spanned by a tower of states $\{\ket{k}\}_{k=0}^{\infty}$.
Similarly as in the case of $\mathfrak{sl}(2)$ spins, one also has to distinguished two distinct realizations of bosonic Fock
spaces $\mathcal{B}^{\pm}$,
\begin{align}
\mathcal{B}^{+}:&\qquad \bb\ket{0} = 0,\quad \bb^{\dagger}\ket{k} = \ket{k+1},\\
\mathcal{B}^{-}:&\qquad \bb^{\dagger}\ket{0} = 0,\quad \bb\ket{k} = \ket{k+1}.
\end{align}
related to each other by an algebra automorphism
\begin{equation}
\mathcal{B}^{+}\to \mathcal{B}^{-}:\qquad \bb^{\dagger}\to \bb,\quad \bb \to - \bb^{\dagger},\quad \mathbf{h} \to -\mathbf{h},
\end{equation}
which is interpreted as the particle-hole conjugation.

By assigning the dissipation to states $I=\{1,2\}$, the off-shell $\Omega$-amplitude is constructed from the Lax operator 
$\bL_{\{1,2\}}(z)$ which carries a representation of algebra
$\mathfrak{A}^{I}_{3,0}\cong \mathfrak{sl}(2)\otimes \mathcal{B}\otimes \mathcal{B}$ and takes the
canonical form of Eq.~\eqref{eqn:canonical_form},
\begin{equation}
\bL_{\{1,2\}}(z) =
\begin{pmatrix}
z + \bJ^{3} - \mathbf{h}_{1} & \bJ^{-} - \bb^{\dagger}_{1}\bb_{2} & \bb^{\dagger}_{1} \\
\bJ^{+} - \bb^{\dagger}_{2}\bb_{1} & z - \bJ^{3} - \mathbf{h}_{2} & \bb^{\dagger}_{2} \\
-\bb_{1} & - \bb_{2} & 1
\end{pmatrix}.
\label{eqn:su3_solution}
\end{equation}
Now is suffices to repeating the logic used before on the $\mathfrak{su}(2)$ case, and define a factorized
off-shell Lax operator $\mathbb{L}^{\omega}_{j}(z)$ for the steady-state solution $\rho_{\infty}$ in the form
\begin{equation}
\mathbb{L}^{\omega}_{j}(z) = \bL^{[1]\,\omega}_{j}(z)\bL^{[2]\,\overline{\omega}}_{-j}(-z).
\end{equation}
This time we equipped each tensor copy with an additional label $\omega$ which, as argued earlier, is needed to supply the
information about the types of $\mathfrak{sl}(2)$ and Fock modules.
After determining the right $\omega$, the coupling constant $g$ is linked to the free representations parameters $z$, $j$
through the boundary equations \eqref{eqn:boundary_conditions} with the solution
\begin{equation}
z = \frac{1}{2},\qquad j = -\frac{\ii}{g},\qquad \omega = (-|-,+).
\end{equation}
The delimiter in $\omega$ was used to explicitly distinguish the $\mathfrak{sl}(2)$ module $\mathcal{V}^{\pm}_{j}$ (on the left) from 
the signatures belonging to the product of Fock spaces $\mathcal{B}^{\pm}$ (on the right, in the ascending order). Specifically,
the above instance requires a lowest-weight type $\mathfrak{sl}(2)$ representation and to assigned $\mathcal{B}^{-}$
($\mathcal{B}^{+}$) to the first (second) bosonic oscillator.

Before heading on to the more involved examples of fermionic models, let us spent a few more words on the non-trivial structure of the 
vacuum $\rvac$ and, in particular, to \emph{inequivalent} roles of the highest and lowest type of (auxiliary) representations.
As said earlier, in order to construct an off-shell $\Omega$-amplitude it is first required to infer an appropriate `internal 
structure' for the auxiliary space of $\bL_{\Lambda_{n+m}}(z)$. Only then it is possible to proceed by solving the corresponding 
finite system of polynomial equations \eqref{eqn:boundary_conditions}.
The upshot here is that the module type labels $\omega$ are essential to assign $\rho_{\infty}$ the appropriate chiral structure.
For instance, an incorrect assignment of the auxiliary bosons which in expression \eqref{eqn:su3_solution},
e.g. by imposing two identical representations $\mathcal{B}^{+}$, would violate the boundary compatibility conditions.
Finally, one can easily verify that the Lax operator $\bL_{\{1,2\}}(z)$ is in fact equivalent to the Lax operator found 
previously in \cite{IP14}.\footnote{In order to exactly match the Lax operator from ref. \cite{IP14} and recover the canonical 
representation of the Lax operator $\bL_{\{1,3\}}$, one should first redefine the algebra generators to eliminate the redundant 
parameter $\eta$, and subsequently apply a diagonal gauge transformation with $G_{L}=1$, $G_{R}={\rm diag}(1,1,-1)$, and ultimately 
the particle-hole transformations on the auxiliary spin and oscillator species.}

\subsection{Fermionic models}
\label{sec:fermionic}

In this section we generalize the above construction to the steady state solutions which pertain to graded $\mathfrak{su}(n|m)$ 
chains, representing the simplest class of interacting integrable models with fermionic degrees of freedom.
We retain the dissipative boundaries given by Eq.~\eqref{eqn:integrable_boundaries}.\footnote{A comment on the Jordan--Wigner 
transformation: When expressed in terms of the fermionic generators, the dissipator attached to the first lattice site differs from 
its non-graded counterpart by a (non-local) Jordan--Wigner `string operator' $W$, indicating that fermionization of the boundary-
driven spin chain maps to a model of non-local dissipation.
This discrepancy between the two formulations which is due to the presence of $W$ is however immaterial as far as only the steady  
states are of our interest, the reason being that $W$ commutes with both the steady state $\rho_{\infty}$ and the total Hamiltonian 
$H^{n|m}$.} Besides bosonic oscillators, the auxiliary particle spaces will now also involve canonical fermions from
two-dimensional spaces $\mathcal{F}$.

The defining $\mathfrak{gl}(1|1)$ representation is spanned by two basis states $\ket{0}$ and $\ket{1}$.
The `highest weight' type representation, denoted by $\mathcal{F}^{+}$, is prescribed by
\begin{equation}
\mathcal{F}^{+}: \qquad \bc^{\dagger} \ket{1} = 0,\quad \bc^{\dagger} \ket{0} = \ket{1},\quad \bc \ket{0} = 0,\quad \bc \ket{1} = \ket{0},
\end{equation}
where the generators obey canonical anticommutation relations
\begin{equation}
\big[\bn,\bc^{\dagger}\big] = \bc^{\dagger},\quad
\big[\bn,\bc\big] = -\bc,\quad
\big[\bc,\bc^{\dagger} \big]=1.
\label{eqn:CAR}
\end{equation}
Similarly, the `lowest weight' representation $\mathcal{F}^{-}$ is obtained from $\mathcal{F}^{+}$ by virtue
of the particle-hole mapping
\begin{equation}
\mathcal{F}^{+}\to \mathcal{F}^{-}:\qquad \bc \to \bc^{\dagger},\quad \bc^{\dagger} \to \bc,\quad \bn \to 1 - \bn.
\end{equation}

\paragraph{Free fermions.}
Arguably the simplest fermionic integrable system is a tight-binding model of \emph{non-interacting} spinless fermions
(with a homogeneous chemical potential) whose interaction is invariant under $\mathfrak{gl}(1|1)$ Lie superalgebra and
in terms of canonical fermions reads\footnote{The interaction $h^{1|1}$ can also be expressed
in terms of fundamental $\mathfrak{su}(2)$ generators. Up to boundary terms this yields the XX model in a homogeneous external field,
that is $h^{1|1}=\sigma^{+}\otimes \sigma^{-}+\sigma^{-}\otimes \sigma^{+}+\tfrac{1}{2}(\sigma^{z}\otimes 1+1\otimes \sigma^{z})$.}
\begin{equation}
h^{1|1}_{i} = c^{\dagger}_{i}c_{i+1} + c^{\dagger}_{i+1}c_{i} - n_{i} - n_{i+1} + 1.
\label{eqn:gl11_interaction}
\end{equation}
In spite of its simplicity, it is remarkable find that the corresponding integrable steady states involve auxiliary spaces with
belong to \emph{non-canonical} $\mathfrak{gl}(1|1)$ representations.

The fermionized integrable reservoirs provided by Eq.~\eqref{eqn:integrable_boundaries} are interpreted as an inflow (outflow) of 
spinless fermions at the left (right) boundary with rate $g$,
\begin{equation}
A_{1}=\sqrt{g}\,E^{21}_{1}\equiv \sqrt{g}\,c^{\dagger}_{1},\qquad A_{N}=\sqrt{g}\,E^{12}_{N}\equiv \sqrt{g}\,c_{N}.
\label{eqn:gl11_driving}
\end{equation}
To find the unique solution to the fixed-point condition \eqref{eqn:fixed_point}, we follows the procedure from the
previous section and first consider the following two partonic Lax elements,
\begin{equation}
\bL_{\{1\}}(z)  =
\begin{pmatrix}
z - (\bn-\tfrac{1}{2}) & \bc^{\dagger} \\
-\bc & 1
\end{pmatrix},\qquad
\bL_{\{2\}}(z) = 
\begin{pmatrix}
1 & \bc \\
\bc^{\dagger} & z + (\bn - \tfrac{1}{2})
\end{pmatrix}.
\end{equation}
By merging them together using the fusion rule (see appendix \ref{app:fusion} for details) we have
\begin{equation}
\bL_{\lambda}(z)\simeq \bL_{\{1\}}^{[1]}(z_{+})\,\bL^{[2]}_{\{2\}}(z_{-}),\qquad
z_{+}=z+\lambda + \tfrac{1}{2},\quad z_{-} = z-\lambda + \tfrac{1}{2}.
\label{eqn:gl11_partonic_fusion}
\end{equation}
The outcome is a two-parameteric Lax operator $\bL_{\lambda}(z)$ whose auxiliary space is identified with
an indecomposable \emph{non-unitary} representation denoted here by $\mathcal{V}_{\lambda}$.
The latter can be realized in terms of canonical generators \eqref{eqn:CAR} as
\begin{equation}
\bL^{\pm}_{\lambda}(z) =
\begin{pmatrix}
z_{+} - \bn & -2\lambda\,\bc^{\dagger} \\
-\bc & z_{-} + \bn
\end{pmatrix},\qquad \lambda = \tfrac{1}{2}(z_{+}-z_{-}),
\end{equation}
where the complex representation parameter $\lambda$ is the \emph{central charge}.\footnote{Verma module 
$\mathcal{V}^{\pm}_{\lambda}$ is a $2$-dimensional indecomposable representation of $\mathfrak{gl}(1|1)$ which is unitary only at 
$\lambda=\tfrac{1}{2}$ (where it coincides with the Fock space representation of canonical fermions). At $\lambda=0$ it becomes 
atypical and reducible.}
To distinguish between the highest- and lowest-weight types of representations we shall use an extra superscript label,
using the convention that $\mathcal{V}^{\pm}_{\lambda}$ is associated with the Fock space $\mathcal{F}^{\pm}$,
i.e. $(\bL^{\pm}_{\lambda})^{12}\ket{0}=0$.

In close analogy to the $\mathfrak{su}(2)$ case, the local unit of the amplitude operator $\Omega_{N}$ is now built from
$\bL$-operator $\bL^{-}_{\lambda}(z)$. The undetermined representation parameters are finally obtained from
the boundary conditions \eqref{eqn:integrable_boundaries}, yielding a unique solution
\begin{equation}
z = \frac{1}{2},\qquad \lambda = \frac{\ii}{g}.
\label{eqn:gl11_Omega_sol}
\end{equation}
The factorization property \eqref{eqn:amplitude_factrorization} implies that $\rho_{\infty}(g)$ is constructed from
a two-component Lax operator $\LL_{\lambda}(z)$ which explicitly reads
\begin{equation}
\mathbb{L}_{\lambda}(z) = \bL^{[1]-}_{\lambda}(z)\,\bL^{[2]+}_{\lambda}(-z),\qquad
{\rm with}\qquad z = \frac{1}{2},\quad \lambda = \frac{\ii}{g}.
\label{eqn:gl11_solution}
\end{equation}
The driving may be reversed by first applying the particle-hole transformation on the physical fermions (see 
Eq.~\eqref{eqn:gl11_driving}), exchanging the order of factors in Eq.~\eqref{eqn:gl11_solution}, and ultimately setting
\begin{equation}
\mathbb{L}_{\lambda}(z) = \bL^{[1]+}_{\lambda}(z)\,\bL^{[2]-}_{\lambda}(-z),\qquad
{\rm with}\qquad z=\frac{1}{2},\quad \lambda=-\frac{\ii}{g}.
\label{eqn:gl11_solution_reversed}
\end{equation}
The auxiliary vacuum state $\rrvac$ remains the product of the lowest- and highest-weight state $\ket{0}$ from the
fermionic Fock modules $\mathcal{F}^{\pm}$.

We find it instructive to remark that the model of free fermions takes a special place among the $\mathfrak{gl}(n|m)$ quantum chains
which plays nicely with the fact that the model is compatible with a larger set of integrable boundary dissipators.
It is rather remarkable however that such an extended set of solutions still admits the Lax representation, albeit the
latter does no longer exhibit the usual additive form. We are not sure whether this enlarged set of steady-state solutions is still
related to representation theory of Yangians. Further details and explicit results are presented in appendix \ref{app:free}.

\paragraph{Example: SUSY t-J model.}
Integrable spin chains whose interactions coincide with graded permutations have been initially
considered in \cite{Kulish86,KS90}. A prominent (and generic) example is the $\mathfrak{su}(1|2)$-symmetric integrable spin chain 
which is mappable to the t-J model at the `supersymmetric point'~\cite{ZR88} ($2t=J$). The spectral problem of the model has been
solved with Bethe Ansatz techniques in \cite{Schlottmann87,Sarkar90,EK92,FK93,GS04}.

The local Hilbert space is now isomorphic to $\mathbb{C}^{1|2}$, and is spanned by a (bosonic) empty state $\ket{0}$ and a pair of
spin-carrying electrons, $\ket{\ua}\equiv c^{\dagger}_{\ua}\ket{0}$ and $\ket{\da}\equiv c^{\dagger}_{\da}\ket{0}$, representing 
fermionic states. The density of interacting can be expanded in terms of canonical fermions and reads ($\sigma = \ua,\da$)
\begin{equation}
h^{1|2}_{i,i+1} =
-\mathcal{P}\Big(c^{\dagger}_{i,\sigma}c_{i+1,\sigma} + c^{\dagger}_{i+1,\sigma}c_{i,\sigma}\Big)\mathcal{P} +
2\big(\vec{S}_{i}\cdot \vec{S}_{i+1}-\tfrac{1}{4}n_{j}n_{i+1}\big)+n_{i}+n_{i+1},
\end{equation}
where $\vec{S}^{\alpha}_{i}=
\tfrac{1}{2}c^{\dagger}_{i,\sigma}\vec{\sigma}_{\sigma,\sigma^{\prime}}c_{i,\sigma^{\prime}}$ and
the projector $\mathcal{P}=\prod_{i=1}^{N}(1-n_{i,\ua}n_{i,\da})$ was used to eliminate the forbidden doubly-occupied state
$\ket{\ua \da}\equiv c^{\dagger}_{\ua}c^{\dagger}_{\da}\ket{0}$.

The grading can be distributed in various ways.\footnote{From the algebraic point of view, distinct inequivalent gradings
indicate that Lie superalgebras do not admit unique simple roots. All distinct possibilities are however related under certain 
boson-fermion duality transformations. In the context of Bethe Ansatz these correspond to inequivalent Bethe vacua and
different ways of proceeding to higher levels in the nesting scheme (see e.g. \cite{GS04,Ragoucy07,Volin12}).}
We shall adopt $|0|=0$, and regard the empty state $\ket{0}$ as the highest-weight state (vacuum) in the physical Hilbert space.
Then, we may consider one of the following three options,
\begin{equation}
(1)\quad
\bigotimes\!\!-\!\!\!-\!\!\bigodot \qquad
(2)\quad
\bigodot\!\!-\!\!\!-\!\!\bigotimes \qquad
(3)\quad
\bigotimes\!\!-\!\!\!-\!\!\bigotimes,
\end{equation}
depicted by the corresponding Kac--Dynkin diagrams.\footnote{By convention we draw an open circle if two
adjacent states are of the same parity and a crossed circle when their parities differ (assuming $|0|=0$), while moving from
left to right.} It is important to remark here that the choice of grading is entirely \emph{independent} from the set
$I=\{\alpha,\beta\}$ which specifies a pair of states subjected to the dissipators.

Let us first set the grading to $|1|=0$ and $|2|=|3|=1$, which corresponds to diagram ($1$).
Incoherent conversion processes induced by the dissipators can be described by any of the following three possibilities:
\begin{equation}
({\rm a})\quad
\ket{0} \longleftrightarrow \ket{\ua} \qquad
({\rm b})\quad
\ket{0} \longleftrightarrow \ket{\da} \qquad
({\rm c})\quad
\ket{\ua} \longleftrightarrow \ket{\da}.
\end{equation}
Options $({\rm a})$ and $({\rm b})$ represent \emph{fermionic} driving and physically corresponds to Markovian transitions between
two states of opposite parities, namely a spin-carrying electron and the unoccupied state $\ket{0}$.
Option $({\rm c})$ is different, and affects the \emph{bosonic} sector (i.e. the $\mathfrak{su}(2)$ doublet) by triggering 
incoherent spin flips.

Let us first address option $({\rm a})$, corresponding to assigning the following pair of Lindblad jump operators
\begin{equation}
A_{1} = \sqrt{g}\,\big(1-n_{1,\ua}\big)c_{1,\da},\qquad
A_{N} = \sqrt{g}\,\big(1-n_{N,\ua}\big)c^{\dagger}_{N,\da}.
\end{equation}
This instance pertains to $I=\{1,2\}$, $\overline{I}=\{3\}$, with $p=q=1$ and $\dot{q}=1$,
which defines the auxiliary algebra $\mathfrak{A}^{\{1,2\}}_{1,2}$ which has the product structure
$\mathfrak{gl}(1|1)\otimes \mathcal{F}\otimes \mathcal{B}$.
The corresponding canonical Lax operator is of the form
\begin{equation}
\bL_{\{1,2\}}(z) =
\begin{pmatrix}
z - \bJ^{11} - \big(\bxid^{13}\bxi^{31} - \tfrac{1}{2}\big) &  \bJ^{12} + \bxid^{13}\bxi^{32} & \bxid^{13} \\
-\bJ^{21} - \bxid^{23}\bxi^{31} & z + \bJ^{22} + \big(\bxid^{23}\bxi^{23} + \tfrac{1}{2}\big) & \bxid^{23} \\
-\bxi^{31} & \bxi^{32} & 1
\end{pmatrix},
\end{equation}
where the generators $\bJ^{ab}$ are associated with the \emph{non-unitary} $\mathfrak{gl}(1|1)$ representation 
$\mathcal{V}_{\lambda}$, whereas the super oscillators are identified with bosonic and fermionic canonical oscillators
in accordance with the rule
\begin{equation}
\bxid^{13} \to \bc^{\dagger},\quad \bxi^{31} \to \bc,\quad
\bxid^{23} \to \bb^{\dagger},\quad \bxi^{32} \to \bb.
\end{equation}
The solution to Eq.~\eqref{eqn:boundary_conditions} is then given in the form
\begin{equation}
\mathbb{L}^{\omega}_{j}(z) = \bL^{[1]\,\omega}_{j}(z)\bL^{[2]\,\overline{\omega}}_{-\overline{j}}(-z),
\label{eqn:gl12_sol2}
\end{equation}
where $\bL^{\omega}_{j}(z)$ now implements the auxiliary algebra
$\mathfrak{A}^{\{1,2\}}_{1,2}=\mathfrak{gl}(1|1)\otimes \mathfrak{osc}(1|1)$, and the representation parameters take the values
\begin{equation}
z = \frac{1}{2},\qquad j = \frac{1}{2} + \frac{\ii}{g},\qquad \omega = (-|-,+).
\end{equation}
In particular, the signature labels $\omega$ (where the bar in Eq.~\eqref{eqn:gl12_sol2} stands for flipping the signs) indicate
that the auxiliary algebra $\mathfrak{A}^{\{1,2\}}_{1,2}$ should be realized in
$\mathcal{V}^{-}_{\lambda}\otimes \mathcal{F}^{-}\otimes \mathcal{B}^{+}$.

The same procedure can be repeated for the case of bosonic driving $({\rm c})$, where the jump operators acts as
\begin{equation}
A_{1} = \sqrt{g}\,c^{\dagger}_{1,\da}c_{1,\ua},\qquad A_{N}=\sqrt{g}\,c^{\dagger}_{i,\ua}c_{i,\da}.
\label{eqn:bosonic_driving}
\end{equation}
The auxiliary algebra $\mathfrak{A}^{\{2,3\}}_{1,2}=\mathfrak{sl}(2)\otimes \mathfrak{osc}(0|2)$ now consists of
a non-compact $\mathfrak{sl}(2)$ module $\mathcal{V}_{j}$ (with the spin generators denoted by $\bJ^{a}$) and a pair
of fermionic Fock spaces $\mathcal{F}\otimes \mathcal{F}$,
\begin{align}
\bL_{\{2,3\}}(z) &=
\begin{pmatrix}
1 & \bc_{1} & \bc_{2} \\
\bc^{\dagger}_{1} & z + \bJ^{3} + \bc^{\dagger}_{1}\bc_{1} - \tfrac{1}{2} & \bJ^{-} + \bc^{\dagger}_{1}\bc_{2} \\
\bc^{\dagger}_{2} & \bJ^{+} + \bc^{\dagger}_{2}\bc_{1} & z + \bJ^{3} + \bc^{\dagger}_{2}\bc_{2} - \tfrac{1}{2}
\end{pmatrix}.
\end{align}
In order to fulfil the boundary constraints, the auxiliary algebra of $\bL^{\omega}_{j}(z)$ should
consist of the product $\mathcal{V}^{-}_{j}\otimes \mathcal{F}^{-}\otimes \mathcal{F}^{+}$.
Finally, $\rho_{\infty}$ is cast in the universal form Eq.~\eqref{eqn:ss_universal}, where now
\begin{equation}
\mathbb{L}^{\omega}_{j}(z) = \bL^{[1]\,\omega}_{j}(z)\bL^{[2]\,\overline{\omega}}_{-j}(-z) \qquad {\rm with} \qquad
z = -\frac{1}{2},\quad j = \frac{\ii}{g},\quad \omega=(-|-,+).
\end{equation}

\section{Vacuum Q-operators}
\label{sec:vacuumQ}

Given that all the solutions are directly related to a particular type of solutions of the graded Yang--Baxter equation \eqref{eqn:graded_YB}, it is quite remarkable (and perhaps surprising) that a one-parametric family of
density matrices $\rho_{\infty}(g)$  \emph{do not} commute for different values of $g$.
On the opposite, an explicit construction of $\rho_{\infty}(g)$ indicates that (reported first in \cite{PIP13})
\begin{equation}
\Big[\rho_{\infty}(g),\rho_{\infty}(g^{\prime})\Big]\neq 0,\qquad {\rm for}\quad g\neq g^{\prime}.
\label{eqn:rho_involution}
\end{equation}
One may still wish argue that due to amplitude factorization \eqref{eqn:amplitude_factrorization} the 
nonequilibrium density operators $\rho_{\infty}(g)$ are not the most `fundamental' physical objects.
Indeed, amplitudes operators $\Omega_{N}(g)$ themselves \emph{do commute} for different values of couplings,
\begin{equation}
\Big[\Omega_{N}(g),\Omega_{N}(g^{\prime})\Big] = 0.
\label{eqn:Omega_involution}
\end{equation}
This shows that $\Omega_{N}(g)$ can be regarded as a family of \emph{vacuum highest-weight transfer matrices}.
However, while the steady states $\rho_{\infty}(g)$ are diagonalizable objects which encode physical properties of the system,
their $\Omega$-amplitudes exhibit a non-trivial \emph{Jordan structure}.\footnote{This is somewhat reminiscent to
what occurs in logarithmic conformal field theories which are governed by non-unitary representation of Virasoro algebra and
possess non-diagonalizable dilatation generators.}
Below we examine this unusual behaviour in more detail and relate it to the vacuum Q-operators.

\paragraph{Baxter's Q-operators.}
Before introducing the notion of vacuum Q-operators, let us first give some comments on the connection between the conventional
Q-operators and Lax operators $\bL_{I}(z)$ introduced in Section~\ref{sec:Yangians}.
The concept of a Q-operator was originally introduced in Baxter's seminal paper on the $8$-vertex model, where it was used
as a device to diagonalize the transfer matrix of the problem by solving a suitable
second-order difference relation~\cite{Baxter72}, and later revived in the context of Potts model~\cite{BS90}
and integrable structure of CFTs~\cite{BLZI,BLZII}. For clarity we focus the subsequent discussion 
entirely on the homogeneous $\mathfrak{su}(2)$ spin chain, providing only a condensed summary of the main ingredients.
For a more comprehensive and pedagogical exposition we refer the reader to \cite{Bazhanov10}.

Baxter's TQ-relation is a functional relation for the fundamental transfer operator $T_{\square}$ of the form
\footnote{For technical reasons we shall think of a closed system and impose twisted boundary conditions.
The case of periodic boundary conditions (i.e. the limit of vanishing twist $\phi\to 0$) exhibits a subtle singular behaviour due to 
restoration of the $SU(2)$ multiplets and has to be treated with care (see~\cite{Bazhanov10}).}
\begin{equation}
T_{\square}(z)Q_{\pm}(z) = T_{0}(z-\tfrac{1}{2})Q_{\pm}(z+1) + T_{0}(z+\tfrac{1}{2})Q_{\pm}(z-1),\quad T_{0}(z) = z^{N}.
\label{eqn:TQ-relations}
\end{equation}
The pair of Baxter Q-operators $Q_{\pm}(z)$ represents two \emph{independent} operator solutions to
the functional equation \eqref{eqn:TQ-relations} and enjoy the involution property
\begin{equation}
\Big[T_{\square}(z),Q_{\pm}(z^{\prime})\Big]=
\Big[Q_{+}(z),Q_{-}(z^{\prime})\Big]=0,\qquad \forall z,z^{\prime}\in \mathbb{C}.
\end{equation}
Eigenvalues of $Q_{\pm}(z)$, denoted by $\mathcal{Q}_{\pm}(z)$, are (up to a twist-dependent phase which is omitted for brevity)
\emph{polynomials} of the form
\begin{equation}
\mathcal{Q}_{-}(z) = \prod_{k=1}^{M}(z-z_{k}),\qquad \mathcal{Q}_{+}(z) = \prod_{k=1}^{N-M}(z-\tilde{z}_{k}).
\end{equation}
Their zeros $z_{k}$ ($\tilde{z}_{k}$) coincide with the Bethe (dual) roots, and are solutions to the celebrated Bethe 
quantization condition
\begin{equation}
\Big(\frac{z+1/2}{z-1/2}\Big)^{N}e^{\pm \ii \phi} = -\frac{\mathcal{Q}_{\mp}(z+1)}{\mathcal{Q}_{\mp}(z+1)}.
\label{eqn:Bethe_equation}
\end{equation}
Polynomiality of eigenvalues of $T_{\square}(z)$ and $Q_{\pm}(z)$ ensure that the TQ-relation \eqref{eqn:TQ-relations} is 
\emph{equivalent} to Bethe equations \eqref{eqn:Bethe_equation}.

Operators $Q_{\pm}$ can be conveniently cast as auxiliary traces over quantum monodromies obtained by the
lattice path integration of partonic Lax operators. Specifically, in the $\mathfrak{su}(2)$ case we have
\begin{equation}
Q_{\pm}(z) \simeq \,{\rm Tr}_{\mathcal{F}}\;
\Big(e^{-\ii \phi\,\bn}\bL_{\pm}(z)\otimes \cdots \otimes \bL_{\pm}(z)\Big),
\label{eqn:Q_su2}
\end{equation}
Here we have made identifications $\bL_{\{1\}}(z)\equiv \bL_{+}(z)$, $\bL_{\{2\}}(z)\equiv \bL_{-}(z)$, and the trace
is with respect to the auxiliary Fock space $\mathcal{F}$.
An analogous construction applies to integrable theories based on higher-rank algebras~\cite{BLZI,BLZII,KLWZ97,KSZ08} where
a \emph{complete} set of Q-operators is associated to Lax operators $\bL_{I}(z)$ introduced in Section \ref{sec:Yangians}.
In the language of Bethe ansatz, this means that eigenvalues of all Q-operators belonging to rational solutions of
Yang--Baxter algebra (cf. Eq.~\eqref{eqn:graded_YB}) are polynomials whose roots coincide with Bethe roots belonging to
different nesting levels. An explicit construction of the full hierarchy of Q-operators for $\mathfrak{gl}(n|m)$ spin 
chains can be found in~\cite{Bazhanov10,Bazhanov11,Frassek11,Frassek13} (and in \cite{KV08,KLT12}, using a different approach).
The outcome of this procedure is a set of $2^{n+m}$ distinct Q-operators which can be arranged on vertices of
a hypercubic lattice~\cite{Tsuboi10}. In this context, partonic Lax operators $\bL_{\{a\}}(z)$ are associated to the
distingusihed set of $n+m$ elementary Q-operators $Q_{\{a\}}(z)$ which can be used to solve the spectral problem by explicit
integrating of an auxiliary linear problem~\cite{KLWZ97,KSZ08}. An important consequence of this is that
eigenvalues of fused transfer matrices which obey the T-system functional identities~\cite{KP92,KNS94} decompose in terms of the 
elementary Q-functions.

\paragraph{Vacuum Q-operators.}
Since in \emph{open} quantum spin chains translational symmetry is manifestly absent, taking (super) traces over auxiliary spaces 
is no longer a priori justified. As originally noticed in \cite{ProsenPRL106}, one may instead consider as a meaningful replacement
projections onto the highest (or lowest) weight states of auxiliary spaces~\footnote{In a more general setting, when particle source 
and drain terms are rotated with respect to the $z$-axis, the highest-or lowest-weight vacua get replaced by spin-coherent 
states~\cite{PKS13}.}. To this end we now define the following set of `vacuum Q-operators'
\begin{equation}
Q^{\rm vac}_{I}(z) = \lvac \bL_{I}(z)\stimes \cdots \stimes \bL_{I}(z) \rvac.
\end{equation}
The previous analysis of the steady-state solutions for $\mathfrak{gl}(n|m)$ spin chains with integrable dissipative
boundaries given by Eq.~\eqref{eqn:integrable_boundaries} shows that all $\Omega$-amplitudes can indeed by identified with vacuum Q-
operators. More specifically, $\Omega$-amplitudes which enter in our nonequilibrium setting always correspond to `mesonic' Lax
operators $\bL_{I}(z)$ with $|I|=2$.

We wish to elaborate on a subtle (but important) point in regard to the auxiliary algebra of $\bL_{I}(z)$ and
the structure of the vacuum state $\rvac$. The fact that $\bL_{I}(z)$ carry (besides Dynkin labels) the information about the types
of irreducible components which implement the auxiliary algebra becomes crucial here.
For instance, already in the simplest case of the $\mathfrak{su}(2)$ chain, we had to define and operate with two
\emph{inequivalent} types of vacuum Q-operators (denoted by $Q^{\rm vac,\pm}_{\{a\}}(z)$, with $a=1,2$) carrying
either of inequivalent bosonic Fock spaces $\mathcal{B}^{\pm}$.
The explicit structure of the fusion relation for partonic operators $\bL_{\{a\}}(z)$
(see appendix \ref{app:fusion}) brings us to the conclusion that the vacuum Q-operators with equal auxiliary modules
are still in involution
\begin{equation}
\Big[Q^{\rm vac,\pm}_{\{a\}}(z),Q^{\rm vac,\pm}_{\{a^{\prime}\}}(z^{\prime})\Big]=0,\qquad \forall z,z^{\prime}\in \mathbb{C},\quad
{\rm and}\quad a,a^{\prime}\in \{1,2\},
\label{eqn:Q_involution}
\end{equation}
and in turn implies that the same property also holds for the corresponding $\Omega$-amplitudes
(as given by Eq.~\eqref{eqn:Omega_involution}).
Conversely, the objects which involve inequivalent auxiliary spaces \emph{do not commute},
\begin{equation}
\Big[Q^{\rm vac,\pm}_{\{a\}}(z),Q^{\rm vac,\mp}_{\{a^{\prime}\}}(z^{\prime})\Big] \neq 0,\qquad \forall z,z^{\prime}\in \mathbb{C},
\quad {\rm and}\quad a,a^{\prime}\in\{1,2\}.
\label{eqn:Q_broken_involution}
\end{equation}
Since the steady-state density operators $\rho_{\infty}(g)$ \emph{always consist of two fused mesonic vacuum Q-operators
of the opposite type} (which is a corollary of property \eqref{eqn:amplitude_factrorization}), by virtue of 
Eq.~\eqref{eqn:Q_broken_involution} they do not inherit the involution property \eqref{eqn:Omega_involution} from
their  amplitude operators $\Omega_{N}(g)$.\footnote{It is instructive to remark that tensor products of irreps of mixed types
do not admit a resolution in terms of a (finite or infinite) discrete sum over extremal-weight irreps, in contrast
to ubiquitous decomposition of tensor products of finite dimensional irreps (or products of extremal-weight irreps of the same type).} 
It remains an interesting open problem to devise a suitable generalization of the Algebraic Bethe Ansatz procedure 
to diagonalize $\rho_{\infty}$~\cite{PIP13}.

\section{Conclusion and outlook}
\label{sec:conclusion}

In this work we introduced a unifying algebraic description for exact nonequilibrium steady states which belong
to an important class of integrable quantum lattice models. We presented an explicit construction of density matrices 
which appear as non-trivial stationary solutions to a non-unitary relaxation process in which a system is coupled to
effective Markovian particle reservoirs attached at its boundaries. We employed a simple set of incoherent particle source and
drain reservoirs which naturally generalized those used previously in refs.~\cite{ProsenPRL107,KPS13,IZ14,PP15,Prosen_review}.
We have shown that such reservoirs partially preserve the integrable structure of the bulk Hamiltonian and permit to obtain
analytic closed-form steady-state density operators in a systematic way.

The solutions were presented in the universal form of a homogeneous fermionic matrix-product operators, and
shown to decompose in terms of the vacuum analogues of Baxter's Q-operators. Such a factorization property reflects the
chiral structure of the states and also allows to reverse directions of particle currents with aid of suitable
particle-hole transformations.
The basic building blocks of our construction are the so-called partonic Lax operators which stem from certain
degenerate representations of graded Yangians, identified recently in~\cite{Bazhanov10,Bazhanov11,Frassek11,Frassek13}.
These rather unconventional algebraic structures admit a canonical realization in terms spins and oscillators.
In the context of our application, these appeared as the auxiliary degrees of freedom in the matrix-product operator representation 
for the steady-state solutions.

The absence of translational symmetry in open quantum chains is profound importance and requires
to replace the usual auxiliary traces by the projectors onto highest- or lowest-weight auxiliary vacua.
The internal algebraic structure of the amplitude operators depends crucially on the parities assigned to particles which
experience dissipation.  In the case of equal parities (bosonic driving), the amplitude operators always involve a single auxiliary 
non-compact $\mathfrak{sl}(2)$ spin, whereas the opposite parities (fermionic driving) require a non-unitary irreducible 
$\mathfrak{gl}(1|1)$ representation which are two dimensional.
The residual auxiliary degrees of freedom pertain to a finite number of canonical (bosonic or fermionic) oscillators which
remain intact upon varying the coupling parameters of the reservoirs.
The universal structure of the steady-state solutions signifies that it is the non-unitary part of the auxiliary algebra which 
ultimately controls their qualitative characteristics: on one end, the presence of $\mathfrak{sl}(2)$ sectors leads to a universal
anomalous (i.e. non-diffusive) $j\sim \mathcal{O}(N^{-2})$ decay of longitudinal currents and cosine-shaped density 
profiles as already found in~\cite{ProsenPRL107,IP14,PP15,Prosen_review}. Fermionic driving is on the other hand
characterized by $\mathfrak{gl}(1|1)$ subspaces and triggers ballistic transport with
non-decaying currents $j\sim \mathcal{O}(N^{0})$ and flat density profiles~\cite{ZnidaricJPA}.
The solutions at hand can therefore be perceived two particular nonequilibrium universality classes.

The distinguished feature of integrable steady states addressed in this work are the non-unitary representations of
Lie (super)algebras . This contrast the conventional approaches to quantum integrable systems
which are primarily based on unitary representations and directly relate to physical excitations in the spectrum
(described by the formalism of Thermodynamic Bethe Ansatz~\cite{YY69,Takahashi71,Gaudin71}).
Physical significance of non-unitary representations in integrable theories is on the
other hand far less understood and has not been much explored in the literature, although a few prominent examples are worth 
mentioning. Most notably, the logarithmic CFTs are based on (non-unitary) reducible indecomposable representations of Virasoro 
algebra~\cite{RS92,Gurarie93} and are known to capture various phenomena in statistical physics ranging from critical dense 
polymers~\cite{PR07}, symplectic fermions~\cite{Kausch00,GRS13}, critical percolation~\cite{Saleur92,Watts96,RP07} to Gaussian 
disordered systems~\cite{CKT96,MS97}. It is perhaps instructive to add that non-compact spin chains are also found in the hadron 
scatting in QCD, which is in the Regge regime governed by the $s=0$ non-compact isotropic Heisenberg magnet~\cite{Lipatov93,FK95}.

The role of non-unitarity in the present nonequilibrium setting is however different as it is not (at least directly) attributed
to physical degrees of freedom, but instead enters on the level of fictitious particles assigned to auxiliary spaces
in a matrix-product representation of nonequilibrium steady states. Nevertheless, it has been found that non-unitary
representations can sometimes be linked to certain hidden conservation laws which turn out to be responsible for anomalous
quantum spin transport (singular DC conductivity) in the linear regime~\cite{ProsenPRL106,IPCMP12,PI13,ProsenPRE14}.

In the conclusion we wish to highlight a few unresolved aspects of the problem which in our opinion deserve to be better explored 
and understood. In order to further extend the range of applicability of the present approach, it is of paramount
importance to obtain better theoretical understanding of the integrability-preserving dissipative boundaries.
In particular, whether there exist a connection between quantum integrability and a special type of Lindblad reservoirs employed here 
remains unanswered at the moment.
Another intriguing open question is to find a field-theoretic version of the Lindbladian evolution which would 
qualitatively reproduce the scaling regime of integrable quantum lattices (cf. \cite{Prosen_review}).
It is moreover difficult to overlook several discernible similarities with the Caldeira--Leggett approach of modelling a
dissipative environment with a boundary-localized friction term~\cite{CL81,CL83}, which has been applied to sine–Gordon theory
with an integrable boundary perturbation~\cite{BLZ99}. In particular, (i) the boundary current is given by the vacuum 
eigenvalues of CFT analogues of Q-operators, (ii) the reservoir parameters are linked to purely imaginary values of highest weights, 
and (iii) the particle current is expressed directly in terms of the nonequilibrium partition function $Z={\rm Tr}\,\varrho_{\infty}$, 
which is otherwise common to both the asymmetric classical exclusion processes~\cite{Blythe07} and their quantum 
counterparts~\cite{Prosen_review} considered here. In our opinion, these curiosities deserve to be further explored in future studies.

\section*{Acknowledgements}
The author thanks P. Claeys, V. Popkov, E. Quinn, and especially T. Prosen for valuable and stimulating discussions and/or
providing comments on the manuscript.

\paragraph{Funding information.}
The author acknowledges support by VENI grant by the Netherlands Organisation for Scientific Research (NWO).

\begin{appendix}

\section{Graded vector spaces and Lie superalgebras}
\label{app:SUSY}

A graded vectors space is a complex vector space $\mathbb{C}^{n|m}$, spanned by basis states $\{\ket{a}\}_{a=1}^{n+m}$,
which is endowed with a $\mathbb{Z}_{2}$ map,
\begin{equation}
p:\qquad \{1,2,\ldots,n+m\}\to \{0,1\},
\label{eqn:grading_map}
\end{equation}
referred to as the \emph{grading}:
\begin{equation}
p(a)\equiv |a| =
\begin{cases}
0 \quad {\rm if}\quad a \; {\rm is}\; {\rm bosonic}\\
1 \quad {\rm if}\quad a \; {\rm is}\; {\rm fermionic}
\end{cases}.
\label{eqn:parity_vectors}
\end{equation}
We subsequently adopt (with no loss of generality) the \emph{distinguished} grading,
\begin{equation}
|a| =
\begin{cases}
0 \quad a\in \{1,2,\ldots,n\}\\
1 \quad a\in \{n+1,\ldots,n+m\}
\end{cases}.
\label{eqn:distinguished_grading}
\end{equation}
This assignment induces a block decomposition on ${\rm End}(\mathbb{C}^{n|m})$, being the space of matrices acting
on $\mathbb{C}^{n|m}$. Specifically, ${\rm End}(\mathbb{C}^{n|m})=\mathcal{V}_{0}\oplus \mathcal{V}_{1}$, 
where components $\mathcal{V}_{0}$ (${\rm dim}\,\mathcal{V}_{0}=n$) and $\mathcal{V}_{1}$ (${\rm dim}\,\mathcal{V}_{1}=m$) represent 
bosonic (even) and fermionic (odd)  parts, respectively. The subspaces $\mathcal{V}_{0}$ and $\mathcal{V}_{1}$ are referred to
as the homogeneous components of ${\rm End}(\mathbb{C}^{n|m})$. Notice that while $\mathcal{V}_{0}$ is a subalgebra,
the odd part $\mathcal{V}_{1}$ is not. A vector space ${\rm End}(\mathbb{C}^{n|m})$ also constitutes
$\mathfrak{gl}(n|m)$ Lie superalgebra. In particular, any element $A$ admits a block form
\begin{equation}
A =
\begin{pmatrix}
A_{00} & A_{01} \\
A_{10} & A_{11}
\end{pmatrix},
\end{equation}
where sub-matrices $A_{00}$, $A_{11}$, $A_{01}$ and $A_{10}$ are of dimensions $n\times n$, $m\times m$,
$n\times m$ and $m\times n$, respectively. The bosonic part decomposes in terms of bosonic subalgebras
$\mathfrak{gl}(n|m)_{0}\cong \mathfrak{gl}(n)\oplus \mathfrak{gl}(m)$ and corresponds to $A_{01}=A_{10}\equiv 0$, whereas
the fermionic (odd) part $\mathfrak{gl}(n|m)_{1}$ pertains to elements with $A_{00}=A_{11}\equiv 0$.

Let $E^{ab}$ denote matrix units, i.e. matrices with the only non-zero element being $1$ in the $a$-th row 
and $b$-th column. Basis element $E^{ab}$ are assigned a $\mathbb{Z}_{2}$-parity according to
$p(E^{ab})\equiv |E^{ab}|\to \{0,1\}$, with the prescription
\begin{equation}
|E^{ab}| = |a| + |b|.
\label{eqn:parity_operators}
\end{equation}
Element $A$ is of even (odd) parity when non-vanishing blocks $A_{ab}$ are of equal (opposite) parity
$|a|+|b|=0$ ($|a|+|b|=1$). For instance, $R$-matrices $R^{n|m}$ are always even elements.
Matrix units $E^{ab}$ form a basis of the fundamental representation of $\mathfrak{gl}(n|m)$ algebra denoted by
$\mathcal{V}^{n|m}_{\square}$. The graded Lie bracket is prescribed by
\begin{equation}
\big[A,B\big] := AB - (-1)^{AB}BA,
\label{eqn:Lie_superbracket}
\end{equation}
and the graded Jacobi identity reads
\begin{equation}
(-1)^{AC}\Big[A,\big[B,C\big]\Big] + (-1)^{AB}\Big[B,\big[C,A\big]\Big] + (-1)^{BC}\Big[C,\big[A,B\big]\Big]=0.
\label{eqn:super_Jacobi}
\end{equation}
Here and below we simplified the notation by writing $(-1)^{a}$ instead of $(-1)^{|a|}$.

Graded vector spaces and Lie superalgebras are a naturally extended over $N$-fold product spaces.
Product spaces inherit the parity according to the prescription
\begin{equation}
|E^{a_{1}b_{1}}\otimes E^{a_{2}b_{2}}\otimes \cdots \otimes E^{a_{N}b_{N}}| = \sum_{k=1}^{N}\big(|a_{k}|+|b_{k}|\big).
\end{equation}
A linear operator $A$ on $(\mathbb{C}^{n|m})^{\otimes N}$ is called a homogeneous element of parity $|A|$ if it satisfies
\begin{equation}
(-1)^{\sum_{k}(a_{k}+b_{k})}A_{a_{1}\ldots a_{N},b_{1}\ldots b_{N}} = (-1)^{A}A_{a_{1}\ldots a_{N},b_{1}\ldots b_{N}}.
\end{equation}
A product of two homogeneous elements $A$ and $B$ has a good parity and is given by $|AB|=|A|+|B|$.
The presence of non-trivial grading also affects the tensor product.
The graded tensor product is denoted by $\stimes$ and defined as
\begin{equation}
A \stimes B = (-1)^{|A|+r(B)}A\otimes B,
\label{eqn:graded_tensor_product}
\end{equation}
where function $r$ designates the row parity,
\begin{equation}
r(E^{a_{1}b_{1}}\otimes E^{a_{2}b_{2}}\otimes \cdots \otimes E^{a_{N}b_{N}}) = \sum_{k=1}^{N}r(a_{k}).
\end{equation}
The advantage of definition \eqref{eqn:graded_tensor_product} is that it preserves the standard tensor multiplication rule,
\begin{equation}
(A \stimes B)(C \stimes D) = AC \stimes BD.
\end{equation}
The graded tensor product can be extended to product spaces by introducing (homogeneous) elements $E^{ab}_{j}$, representing
the generators associated to the $j$-th copy of ${\rm End}(\mathbb{C}^{n|m})$. Notice that in contrast to the standard
(non-graded) basis $1\otimes \cdots \otimes E^{ab}\otimes \cdots \otimes 1$ of ${\rm End}(\mathbb{C}^{n+m})^{\otimes N}$,
elements $E^{ab}_{j}$ do not commute at different lattice sites, but we find instead
\begin{equation}
E^{ab}_{i}E^{cd}_{j}=(-1)^{(a+b)(c+d)}E^{cd}_{j}E^{ab}_{i}.
\label{eqn:graded_commutation}
\end{equation}
On the same lattice site they however still obey the property of projectors,
\begin{equation}
E^{ab}_{i}E^{cd}_{i}=\delta_{cb}\,E^{ad}_{i}.
\label{eqn:graded_projector}
\end{equation}
The last two properties combined yield
\begin{equation}
\Big[E^{ab}_{j},E^{cd}_{k}\Big] = \delta_{jk}\Big(\delta_{cb}\,E^{ad}_{k}-
(-1)^{(a+b)(c+d)}\delta_{ad}\,E^{cb}_{j}\Big).
\end{equation}
The graded generators acting on the $N$-particle space $(\mathbb{C}^{n|m})^{\otimes N}$ read in terms of the graded tensor product
\begin{equation}
E^{ab}_{i} = 1^{\otimes (i-1)} \stimes E^{ab} \stimes 1^{\otimes (N-i)},
\end{equation}
whereas expressed in terms of the standard tensor product they assume the expansion
\begin{equation}
E^{ab}_{i} = (-1)^{(a+b)\sum_{k=j+1}^{N}c_{k}} 1^{\otimes (i-1)}\otimes E^{ab}\otimes E^{c_{j+1}c_{j+1}}\otimes
\cdots \otimes E^{c_{N}c_{N}}.
\label{eqn:general_JW}
\end{equation}
This prescription should be interpreted as the higher-rank version of the Jordan--Wigner transformation~\cite{GM98}.

Interaction densities $h^{n|m}$ for a class of the so-called `fundamental graded models' are identified with
graded permutations $P^{n|m}$ on $\mathbb{C}^{n|m}\otimes \mathbb{C}^{n|m}$,
\begin{equation}
P^{n|m} = (-1)^{b}E^{ab}\stimes E^{ba}.
\end{equation}
Permutations $P^{n|m}$ can be alternatively given also as matrices acting on the two-fold fundamental spaces
$\mathbb{C}^{n+m}\otimes \mathbb{C}^{n+m} $, reading
\begin{equation}
P^{n|m} = (-1)^{a+b}E^{ab}\otimes E^{ba}.
\end{equation}

The defining $\mathfrak{su}(n|m)$ representations admit realizations in terms of canonical fermions.
In the $\mathfrak{su}(1|1)$ case, the graded projectors $\calE_{i}$ act non-identically only on the $i$-th copy
of $\mathbb{C}^{1|1}$ in the chain, and are realized as a $2\times 2$ matrix of spinless fermions
\begin{equation}
\calE_{i} = 
\begin{pmatrix}
1-n_{i} & c_{i} \\
c^{\dagger}_{i} & n_{i}
\end{pmatrix}.
\label{eqn:su11_projector}
\end{equation}
Here the generators $n_{i}$ and $1-n_{i}$ span the even (bosonic) subalgebra $\mathcal{V}_{0}$, while $c_{i}$ and $c^{\dagger}_{i}$
are the fermionic generators which span the odd part $\mathcal{V}_{1}$ and satisfy canonical anticommutation relations
\begin{equation}
\{c_{i},c^{\dagger}_{j}\} = \delta_{j,k},\quad
\{c_{i},c_{j}\}=\{c^{\dagger}_{i},c^{\dagger}_{j}\}=0.
\end{equation}
Equation \eqref{eqn:general_JW} is nothing but the well-known Jordan--Wigner transformation from Pauli 
spins to canonical spinless fermions
\begin{equation}
\begin{split}
c^{\dagger}_{i} = 1^{\otimes (i-1)}\otimes \sigma^{-} \otimes (\sigma^{z})^{\otimes (N-1)},\\
c_{i} = 1^{\otimes (i-1)}\otimes \sigma^{+} \otimes (\sigma^{z})^{\otimes (N-1)}.
\end{split}
\end{equation}

In systems with multiple fermionic species (e.g. spin-carrying fermions), the super projectors can be constructed
with aid of the fusion procedure~\cite{GM98}. For instance, the local physical space of a $\mathfrak{su}(2|2)$ spin chain is four 
dimensional, spanned by states
\begin{equation}
\ket{0},\quad \ket{\ua}\equiv c^{\dagger}_{\ua}\ket{0},\quad
\ket{\da}\equiv c^{\dagger}_{\da}\ket{0},\quad \ket{\ua \da}\equiv c^{\dagger}_{\ua}c^{\dagger}_{\da}\ket{0}.
\end{equation}
At each lattice site $i$ we thus have $4\times 4=16$ generators,
\begin{equation}
\calE^{ac,bd}_{\ua \da} = (\calE_{\ua} \stimes \calE_{\da})^{ac,bd} \equiv
(-1)^{|a+b||c|}\calE^{ab}_{\ua}\calE^{cd}_{\da}.
\end{equation}
Flattening the indices readily yields the graded permutation on $\mathbb{C}^{2|2}\otimes \mathbb{C}^{2|2}$, taking the form
\begin{equation}
P^{2|2} = (-1)^{b}\calE^{ab}_{\ua \da}\stimes \calE^{ba}_{\ua \da}.
\end{equation}
Furthermore, the fermionic realization of the graded projector $P^{2|1}$ can be obtained from $P^{2|2}$ by projecting out
e.g. the doubly-occupied state $\ket{\ua \da}$. The local space of states thus consists of the triplet
\begin{equation}
\ket{0},\quad c^{\dagger}_{\ua}\ket{0},\quad c^{\dagger}_{\da}\ket{0}.
\end{equation}
Choosing e.g. the grading as $|0|=0$ and $|1|=|2|=1$, one finds
\begin{equation}
\calE_{i} =
\begin{pmatrix}
(1-n_{i,\ua})(1-n_{i,\da}) & (1-n_{i,\ua})c_{i,\da} & c_{i,\ua}(1-n_{i,\da}) \\
(1-n_{i,\ua})c^{\dagger}_{i,\da} & (1-n_{i,\ua})n_{i,\da} & c^{\dagger}_{i,\da}c_{i,\ua} \\
c^{\dagger}_{i,\ua}(1-n_{i,\da}) & c^{\dagger}_{i,\ua}c_{i,\da} & n_{i,\ua}(1-n_{i,\da})
\end{pmatrix}.
\end{equation}

\section{Fusion and factorization properties of Lax operators}
\label{app:fusion}

In this section we revisit the main factorization and fusion formulae for the $\mathfrak{gl}(n|m)$ integrable spin chains.
A comprehensive and more detailed can be found e.g. in references~\cite{Bazhanov10,Bazhanov11,Frassek11,Frassek13}.

\paragraph{Classification of irreducible highest-weight $\mathfrak{gl}(n|m)$ modules.}
Before presenting the fundamental factorization property of the rational Lax operators we need to introduce
irreducible highest-weight $\mathfrak{gl}(n|m)$ representations. These are known as the \emph{Verma modules},
denoted by $\mathcal{V}^{+}_{\Lambda_{n+m}}$, and are characterized by (i) the highest-weight property
\begin{equation}
\bJ^{a,a+1}\ket{\rm hws} = 0 \qquad {\rm for}\qquad a=1,2,\ldots, n+m-1,
\end{equation}
and (ii) the weight vector $\Lambda_{n+m}=(\lambda_{1},\ldots,\lambda_{n},\lambda_{n+1},\ldots,\lambda_{n+m})$ through
the action of Cartan generators (no summation over repeated indices),
\begin{equation}
\bJ^{aa} \ket{\rm hws} = \lambda_{a}\ket{\rm hws}.
\end{equation}
A representation $\Lambda_{n}$ is typically of \emph{infinite} dimension, corresponding to generic complex-valued
weights $\lambda_{a}$. Since we shall mostly need the restriction to $\mathfrak{sl}(n|m)$ subalgebra, we introduce
the $\mathfrak{sl}(n|m)$ weights as
\begin{equation}
\mu_{a} = (-1)^{a}\lambda_{a}-(-1)^{a+1}\lambda_{a+1},\qquad a=1,2,\ldots, n-1.
\end{equation}
In the case of unitary $\mathfrak{sl}(n|m)$ representations, all $\mu_{a}$ for $a\neq n$ must be non-negative integers, while
$\mu_{n}$ can take arbitrary real values. The fundamental representation of $\mathfrak{sl}(n|m)$ is given by the weight vector 
$\Lambda_{n+m}=(1,0,\ldots,0)$. Kac--Dynkin labels and finite-dimensional irreducible representations are in one-to-one 
correspondence, with Young diagrams corresponding to non-negative non-increasing weights. Rectangular representations $\{s,a\}$ with
$s$ columns and $a$ rows have a single non-vanishing label $\mu_{a}=s$.

\subsection{Factorization of Lax operators}

Highest-weight $\mathfrak{gl}(n|m)$-invariant transfer operators $T^{+}_{\Lambda_{n+m}}(z)$
acting on a Hilbert space $\mathcal{H}\cong (\mathbb{C}^{n|m})^{\otimes N}$ are given by\footnote{Here it is implicitly assumed that
the super trace exists. Additional regulators in the form of boundary twists may be needed in general.}
\begin{equation}
T^{+}_{\Lambda_{n+m}}(z) = {\rm Str}_{\rm \mathcal{V}^{+}_{\Lambda_{n+m}}}
\underbrace{\bL_{\Lambda_{n+m}}(z)\stimes \cdots \stimes \bL_{\Lambda_{n+m}}(z)}_{N\;{\rm copies}},
\end{equation}
which due to Yang--Baxter relation \eqref{eqn:graded_YB} enjoy the commutativity property,
\begin{equation}
\big[T^{+}_{\Lambda_{n+m}}(z),T^{+}_{\Lambda^{\prime}_{n+m}}(z^{\prime})\big]=0.
\end{equation}
for all $z,z^{\prime}\in \mathbb{C}$ and representation labels $\Lambda_{n+m}$ and $\Lambda^{\prime}_{n+m}$.
Remarkably however, operators $T^{+}_{\Lambda_{n+m}}$ do not represent the most elementary objects in the theory.
In fact, they factorize\footnote{The algebraic origin of the factorization formula has to do with the
$\mathcal{U}(\mathfrak{g})$-invariant universal $\mathcal{R}$-matrix which decomposes in terms of tensor products of components from 
the corresponding Borel subalgebras~\cite{BLZI,BLZII}.} into an ordered sequence of Q-operators $Q_{\{a\}}$\footnote{Geometrically,
Q-operators can understood as  Pl\"{u}cker coordinates on Grassmannian manifolds~\cite{KLWZ97,Tsuboi13}.},
\begin{equation}
T^{+}_{\Lambda_{n+m}}(z) \simeq Q_{\{1\}}(z_{1}+\lambda^{\prime}_{1})Q_{\{2\}}(z_{2}+\lambda^{\prime}_{2})\cdots
Q_{\{n+m\}}(z_{n+m}+\lambda^{\prime}_{n+m}).
\label{eqn:T_factorization}
\end{equation}
Here parameters $\lambda^{\prime}_{a}$ are the `shifted weights',
\begin{equation}
\Lambda^{\prime}_{n+m}=\Lambda_{n+m}+\varrho_{n+m},\qquad
\varrho_{a} = \sum_{b=a+1}^{n+m}\tfrac{1}{2}(-1)^{b}-\sum_{b=1}^{a-1}\tfrac{1}{2}(-1)^{b}.
\end{equation}
In the $\mathfrak{gl}(n)$ case, i.e. when $m=0$, the shifts arrange in `complete $n$-strings', reading
$\varrho_{n}=\tfrac{1}{2}(n-1,n-3,\ldots 1-n)$, closely resembling the pattern of the string-type solutions to 
Bethe Ansatz equations. Indeed, factorization property \eqref{eqn:T_factorization} is a direct consequence of the local
factorization relation~\cite{Bazhanov11}
\begin{equation}
\bL_{\{1\}}(z_{1}+\lambda^{\prime}_{1})\bL_{\{2\}}(z_{2}+\lambda^{\prime}_{2})\cdots \bL_{\{n+m\}}(z_{n+m}+\lambda^{\prime}_{n+m})=
\bS\,\bL^{+}_{\Lambda_{n+m}}(z)\,\bK\,\bS^{-1},
\end{equation}
Below we exemplify the factorization procedure on a few concrete instances.

\paragraph{Basic example: $\mathfrak{sl}(2)$ case.}
The factorization property is best illustrated on the $\mathfrak{sl}(2)$ case. The corresponding highest-weight Lax operators read
\begin{equation}
\bL^{+}_{\Lambda_{2}}(z) =
\begin{pmatrix}
z + \bJ^{3} & \bJ^{-} \\
\bJ^{+} & z - \bJ^{3}
\end{pmatrix},
\label{eqn:sl2_Lax}
\end{equation}
and are characterized by a single Dynkin label $j$ which parametrized the $\mathfrak{gl}(2)$ weight vector $\Lambda_{2}=(j,-j)$.
The non-compact spin generators $\bJ^{a}$ act on a $\mathfrak{sl}(2)$ module $\mathcal{V}^{+}_{j}$,
and can be conveniently given in the Holstein--Primakoff form
\begin{equation}
\bJ^{3} = j - \bb^{\dagger} \bb,\quad \bJ^{+} = \bb,\quad \bJ^{-} = \bb^{\dagger}(2j-\bb^{\dagger} \bb),
\label{eqn:HP}
\end{equation}
where $\bb$ and $\bb^{\dagger}$ are the generators of a bosonic oscillator obeying canonical commutation relations
\begin{equation}
\big[\bb,\bb^{\dagger}\big] = 1,\quad
\big[\mathbf{h},\bb \big] = -\bb,\quad
\big[\mathbf{h},\bb^{\dagger} \big] = \bb^{\dagger},
\label{eqn:canonical_bosonic_relations}
\end{equation}
and $\mathbf{h}=\bb^{\dagger} \bb + \tfrac{1}{2}$ is the mode number operator.
We furthermore define two types of Fock space representations, denoted by
\begin{align}
\begin{split}
\mathcal{B}^{+}:&\qquad \bb\ket{0} = 0,\quad \bb^{\dagger}\ket{k}=\ket{k+1}, \\
\mathcal{B}^{-}:&\qquad \bb^{\dagger}\ket{0} = 0,\quad \bb\ket{k}=\ket{k+1}.
\label{eqn:bosonic_Fock_spaces}
\end{split}
\end{align}
The two are related to each other under the particle-hole transformation
$\bb \to \bb^{\dagger}$, $\bb^{\dagger}\to \bb$ and $\mathbf{h} \to -\mathbf{h}$.

A pair of partonic Lax operators $\bL_{\{1\}}(z)$ and $\bL_{\{2\}}(z)$ can be straightforwardly obtained from $T^{+}_{\Lambda_{2}}(z)$
by considering two possible way of taking a (correlated) large-$j$ and large-$z$ limits (cf. \cite{Bazhanov10}). This is achieved by 
keeping either of the combinations $z_{\pm}=z\pm(j+\tfrac{1}{2})$ fixed, resulting in the `degenerate' Lax operators of the form
\begin{equation}
\bL_{\{1\}}(z) =
\begin{pmatrix}
z - \mathbf{h} & \bb^{\dagger} \\
-\bb & 1
\end{pmatrix},\qquad
\bL_{\{2\}}(z) = 
\begin{pmatrix}
1 & \bb^{\dagger} \\
-\bb & z + \mathbf{h}
\end{pmatrix},
\label{eqn:sl2_partonic}
\end{equation}
which represent two distinct well-defined and valid solutions to the Yang--Baxter equation.

\subsection{Fusion of partonic Lax operators}
Quantum groups are endowed with a coproduct, ensuring that the algebraic structure gets preserved under
tensor multiplication $\mathcal{Y}\to \mathcal{Y}\stimes \mathcal{Y}$. Partonic Lax operators, as defined in Eq.~\eqref{eqn:partonic},
represent the simplest solutions of the Yang--Baxter equation \eqref{eqn:graded_YB}. They serve as irreducible components for
obtaining other realizations of $\mathcal{Y}$ via fusion. Below we outline the main features of such a procedure,
while referring the reading for a more complete and detailed presentation to references \cite{Bazhanov11,Frassek11,Frassek13}.

Let $I,J\subseteq \{1,2,\ldots,n+m\}$ be two index sets. We shall only consider operators $\bL(z)$ which are linear in spectral 
parameter $z$, requiring the sets $I$ and $J$ to be non-intersecting, $I\cap J = \emptyset$.
Set $I$ (resp. $J$) comprises of $p$ ($\dot{p}$) bosonic and $q$ ($\dot{q}$) fermionic indices.
We furthermore introduce $K\equiv I\cup J$, involving $\ddot{p}$ ($\ddot{q}$) bosonic (fermionic) indices, such that 
$p+\dot{p}+\ddot{p}=n$ and $q+\dot{q}+\ddot{q}=m$.
Fusion is a process of merging two canonical Lax operators $\bL_{I}$ and $\bL_{J}$ of respective ranks
$|I|=p+q$ and $|J|=\dot{p}+\dot{q}$, which takes the abstract form
\begin{equation}
\bL_{K}(z) \sim \bL^{[1]}_{I}(z+z_{1})\,\bL^{[2]}_{J}(z+z_{2}),
\label{eqn:fusion_abstract}
\end{equation}
for some appropriate choice of shifts $z_{1}$ and $z_{2}$. The superscript square brackets were needed
here to distinguish inequivalent species. The precise prescription for the fusion rule is entailed by the following form
\begin{equation}
\bL^{[1]}_{I}\Big(z+\tfrac{1}{2}\sum_{\dot{d}\in J}(-1)^{\dot{d}}\Big)\,
\bL^{[2]}_{J}\Big(z-\lambda - \tfrac{1}{2}\sum_{d\in I}(-1)^{d}\Big) = \bS\,\bL^{[1]}_{K}(z)\,\bK^{[2]}\,\bS^{-1},
\label{eqn:fusion_rule}
\end{equation}
where $\bK$ is a triangular  `disentangling matrix' and $\bS$ a suitable global similarity transformation.
An implication of fusion formula \eqref{eqn:fusion_rule} is that Lax operators which are realized in terms of algebras
$\mathfrak{A}^{K}_{m,n}$ are not elementary, but instead factorize according to
$\mathfrak{A}^{K}_{m,n}\to \mathfrak{A}^{I}_{m,n}\otimes \mathfrak{A}^{J}_{m,n}$.
Below we take a closer look at this by inspecting a few explicit examples.

We begin by noticing that the above fusion procedure clearly violates the canonical form given by Eq.~\eqref{eqn:canonical_form},
as it appears to involve an exceeding number of auxiliary spaces. We shall in turn demonstrate that all redundant auxiliary
spaces can be eliminated upon appropriately redefining the generators.
In particular, there exist a canonical procedure to reduce the number of oscillators
from $|I|\cdot |\overline{I}|+|J|\cdot |\overline{J}|$ down to $|K|\cdot |\overline{K}|$ and expressing
the $\mathfrak{gl}(|K|)$  generators in terms of independent  generators of $\mathfrak{gl}(|I|)$ and $\mathfrak{gl}(|J|)$,
dressed with $|K|$ additional oscillators. This is in practice achieved by virtue of the homomorphisms~\cite{Frassek11}
\begin{equation}
\mathfrak{gl}(|K|)\to \mathfrak{gl}(p|q)\otimes \mathfrak{gl}(\dot{p}|\dot{q})\otimes \mathfrak{osc}(p+\dot{p}|q+\dot{q}),
\label{eqn:homomorphisms}
\end{equation}
in which the post-fusion $\mathfrak{gl}(p+\dot{p}|q+\dot{q})$ generators $\widehat{\bJ}^{ab}$ are given by the following prescription
\begin{equation}
\begin{split}
\widehat{\bJ}^{ab} &= \bJ^{ab}_{1} + \bxid^{a\dot{c}}_{1}\bxi^{\dot{c}b}_{1}, \\
\widehat{\bJ}^{a\dot{b}} &= \lambda(-1)^{\dot{b}}\bxid^{a\dot{b}}_{1} -
(-1)^{(\dot{b}+\dot{d})(\dot{b}+c)}\bxid^{a\dot{d}}_{1}\bxid^{c\dot{b}}_{1}\bxi^{\dot{d}c}_{1}+
\bxid^{a\dot{c}}_{1}\bJ^{\dot{c}\dot{b}}_{2} - (-1)^{\dot{b}+c}\bJ^{ac}_{1}\bxid^{c\dot{b}}_{1}, \\
\widehat{\bJ}^{\dot{a}b} &= \bxi^{\dot{a}b}_{1}, \\
\widehat{\bJ}^{\dot{a}\dot{b}} &= \bJ^{\dot{a}\dot{b}}_{2} + \lambda (-1)^{\dot{b}}\delta_{\dot{a}\dot{b}}-
(-1)^{(\dot{a}+\dot{b})(\dot{b}+c)}\bxid^{c\dot{b}}_{1}\bxi^{\dot{a}c}_{1}.
\end{split}
\label{eqn:oscillator_realization}
\end{equation}
with $\bJ^{ab}_{1}$ and $\bJ^{\dot{a}\dot{b}}_{2}$ denoting the $\mathfrak{gl}(p|q)$ and $\mathfrak{gl}(\dot{p}|\dot{q})$ super
spins, respectively, whereas the oscillators are to be identified as
\begin{equation}
\bxi_{\dot{a}b} =
\begin{cases} \bxi^{[1]}_{\dot{a}b}, & b \in I \\
\bxi^{[2]}_{\dot{a}b}, & b \in J
\end{cases},\qquad
\bxid_{a\dot{b}} =
\begin{cases} \bxid^{[1]}_{a\dot{b}}, & a \in I \\
\bxid^{[2]}_{a\dot{b}}, & a \in J
\end{cases}.
\end{equation}
The latter are either bosonic or fermionic, depending on the grading.
Finally, the tridiagonal matrix $\bK$ is of the form
\begin{equation}
\bK =
\begin{pmatrix}
1 & -(-1)^{\dot{b}}\bxi^{a\dot{b}}_{2} & 0 \\
0 & 1 & 0 \\
0 & 0 & 1
\end{pmatrix},
\label{eqn:general_K}
\end{equation}
while the similarity transformation in Eq.~\eqref{eqn:fusion_rule} reads
\begin{equation}
\bS= \exp{\left(\sum_{a\in I}\sum_{\dot{b}\in J}\sum_{\ddot{c}\in \overline{K}}
\bxid^{a\dot{b}}_{1}\big((-1)^{a}\bxid^{\dot{b}a}_{2}+\bxid^{\dot{b}\ddot{c}}_{2}\bxi^{\ddot{c}a}_{1}\big)\right)},
\label{eqn:general_S}
\end{equation}
where the double-dotted indices represent the summation over $K$.

\paragraph{$\mathfrak{sl}(2)$ case.}
The basic principle of fusion can be explained on the $\mathfrak{sl}(2)$ theory. Fusion can be understood as
the opposite procedure of factorization which is outlined in the previous section.
The partonic pieces given by expressions \eqref{eqn:sl2_partonic} can be fused in two distinct ways.
To this end we introduce square brackets and assign a boson oscillator to each tensor factor, yielding
the following operator identity on $\mathbb{C}^{2}\otimes \mathcal{B}\otimes \mathcal{B}$,
\begin{align}
\bL^{[1]}_{\{2\}}(z_{2})\bL^{[2]}_{\{1\}}(z_{1}) &=
\begin{pmatrix}
1 & \bb^{\dagger}_{1} \\
\bb_{1} & z_{2} + \bn_{1}
\end{pmatrix}
\begin{pmatrix}
z_{1} - \bn_{2} & \bb^{\dagger}_{2} \\
-\bb_{2} & 1
\end{pmatrix} \nonumber \\
&= \exp{\big(\bb^{\dagger}_{1} \bb_{2}\big)}
\begin{pmatrix}
1 & 0 \\
\bb_{1} & 1
\end{pmatrix}
\begin{pmatrix}
z + \bJ^{3}_{2} & \bJ^{-}_{2} \\
\bJ^{+}_{2} & z - \bJ^{3}_{2}
\end{pmatrix}
\exp{\big(-\bb^{\dagger}_{1}\bb_{2}\big)},
\label{eqn:su2_fusion_rule}
\end{align}
where the input spectral parameters are given by
\begin{equation}
z_{1}=z+j+\tfrac{1}{2},\qquad z_{2}=z-j-\tfrac{1}{2}.
\label{eqn:fusion_parameter_constraints}
\end{equation}
Notice that in the second line of Eq.~\eqref{eqn:su2_fusion_rule} the oscillators have been rearranged
using the similarity transformation
$\bS=\exp{(\bb^{\dagger}_{1}\bb_{2})}$ on $\mathcal{B}\otimes \mathcal{B}$, which reads explicitly
\begin{equation}
\bS\,\bb^{\dagger}_{1}\,\bS^{-1} = \bb^{\dagger}_{1},\quad
\bS\,\bb_{1}\,\bS^{-1} = \bb_{1} + \bb_{2},\quad
\bS\,\bb^{\dagger}_{2}\,\bS^{-1} = \bb^{\dagger}_{1}+\bb^{\dagger}_{2},\quad
\bS\,\bb_{2}\,\bS^{-1} = \bb_{2}.
\end{equation}
On the other hand, fusing in the opposite order yields a similar operator identity
\begin{equation}
\bL^{[1]}_{\{1\}}(z_{+})\bL^{[2]}_{\{2\}}(z_{-}) =
\exp{\big(\bb^{\dagger}_{1}\bb^{\dagger}_{2}\big)}
\begin{pmatrix}
z + \bJ^{3}_{1} & \bJ^{-}_{1} \\
\bJ^{+}_{1} & z - \bJ^{3}_{1}
\end{pmatrix}
\begin{pmatrix}
1 & -\bb_{2} \\
0 & 1
\end{pmatrix}
\exp{\big(-\bb^{\dagger}_{1}\bb^{\dagger}_{2}\big)},
\end{equation}
where again the parameter constraints \eqref{eqn:fusion_parameter_constraints} are imposed.

Formula \eqref{eqn:su2_fusion_rule} readily implies factorization property for the highest-weight $\mathfrak{sl}(2)$-invariant
transfer operator $T^{+}_{\Lambda_{2}}(z)\equiv T^{+}_{j}(z)$ in term of a pair of Q-operators,
\begin{equation}
T^{+}_{j}(z) = Q_{\{1\}}(z+j+\tfrac{1}{2})Q_{\{2\}}(z-j-\tfrac{1}{2}).
\end{equation}
A sequence of transfer matrices $T_{j}(z)$ with $2j \in \mathbb{Z}$, pertaining to finite-dimensional irreducible $\mathfrak{su}(2)$ 
representations, can the be obtained from $T^{+}_{j}(z)$ with aid of the Bernstein--Gelfand--Gelfand resolution of
finite-dimensional modules, $\mathcal{V}_{j} = \mathcal{V}^{+}_{j}-\mathcal{V}^{+}_{-j-1}$, resulting in
Bazhanov--Reshetikhin determinant representation~\cite{BR90}
\begin{equation}
T_{j}(z) = Q_{\{1\}}(z+j+\tfrac{1}{2})Q_{\{2\}}(z-j-\tfrac{1}{2}) - Q_{\{2\}}(z+j+\tfrac{1}{2})Q_{\{1\}}(z-j-\tfrac{1}{2}).
\end{equation}

A comment in regard to the so-called vacuum Q-operators is in order here. First, recall that vacuum Q-operators represent
a family of transfer operators which are constructed from a path-ordered product of (partonic) Lax operators contracted with 
respect to a suitable `vacuum state'. Vacuum state are presently identified with the highest (or lowest) weight state
in $\mathcal{B}$.
Mutual commutativity of vacuum Q-operators can be inferred from fusion formula \eqref{eqn:su2_fusion_rule}, after
observing that (i)  when the two Fock space involved are of same type the product vacuum $\ket{0}\otimes \ket{0}$ remains inert
under the action of the similarity transformation $\bS$, i.e. $\bS \ket{0}\otimes \ket{0}=\ket{0}\otimes \ket{0}$, and (ii)
the disentangler $\bK$ has no global effect due to its triangular form.
In the opposite scenario, a fusion of two Lax operators which involve two different types of Fock spaces inevitably excites
the vacuum to a coherent state which in turn prevents the vacuum T-operator from decomposing into two vacuum Q-operators.
For the very same reason vacuum Q-operators pertaining to inequivalent auxiliary modules are not guaranteed to satisfy
the involution property. In fact, it can be explicitly confirmed that they do not.

\paragraph{$\mathfrak{sl}(3)$ case.}
The $\Omega$-amplitudes for the integrable steady states constructed in this work are all formed from
`mesonic' Lax operators, namely objects which result from the fusion of two partonic elements.
As an explicit example we consider the $SU(3)$-symmetric Lai--Sutherland chain~\cite{Lai74,Sutherland75}, to which we
ascribe auxiliary algebra $\mathfrak{A}^{\{1\}}_{3,0}\otimes \mathfrak{A}^{\{2\}}_{3,0} \to \mathfrak{A}^{\{1,2\}}_{3,0}$.
Setting $I=\{1\}$ and $J=\{2\}$, the fusion formula is of the form (using the shifted weights $\varrho_{2}=\tfrac{1}{2}(1,-1)$)
\begin{equation}
\bL^{[1]}_{\{1\}}(z+\lambda+\tfrac{1}{2})\bL^{[2]}_{\{2\}}(z-\tfrac{1}{2}) = \bS\,\bL^{[1]}_{\{1,2\}}(z)\,\bK^{[2]}\,\bS^{-1},
\end{equation}
with the partonic Lax operators reading
\begin{align}
\bL^{[1]}_{\{1\}}(z) &=
\begin{pmatrix}
z + j_{1} - \bh^{[1]}_{1} - \bh^{[1]}_{2} & \bb^{[1]\dagger}_{1} & \bb^{[1]\dagger}_{2} \\
-\bb^{[1]}_{1} & 1 & 0 \\
-\bb^{[1]}_{2} & 0 & 1
\end{pmatrix},\\
\bL^{[2]}_{\{2\}}(z) &=
\begin{pmatrix}
1 & -\bb^{[2]}_{1} & 0 \\
\bb^{[2]\dagger}_{1} & z + j_{2} -\bh^{[2]}_{1} - \bh^{[2]}_{2} & \bb^{[2]\dagger}_{2} \\
0 & -\bb^{[2]}_{2} & 1
\end{pmatrix}.
\end{align}
The oscillators are disentangled with aid of
\begin{equation}
\bK^{[2]} =
\begin{pmatrix}
1 & -\bb^{[2]}_{1} & 0 \\
0 & 1 & 0 \\
0 & 0 & 1
\end{pmatrix},
\end{equation}
and an additional similarity transformation $\bS=\mathbf{U}\,\mathbf{V}$, with
\begin{align}
\mathbf{U} = \exp{\Big(\bb^{[1]\dagger}_{1}\bb^{[2]\dagger}_{1}\Big)},\quad
\mathbf{V} = \exp{\Big(\bb^{[1]\dagger}_{1}\bb^{[2]\dagger}_{2}\bb^{[1]}_{2}\Big)}.
\end{align}
Explicitly, these transformations act as
\begin{equation}
\begin{split}
\mathbf{U}\, \bb^{[1]}_{1}\, \mathbf{U}^{-1} = \bb^{[1]}_{1} - \bb^{[2]\dagger}_{1},\\
\mathbf{U}\, \bb^{[2]}_{1}\, \mathbf{U} ^{-1} = \bb^{[2]}_{1} - \bb^{[1]\dagger}_{1},
\end{split}
\end{equation}
and
\begin{equation}
\begin{split}
\mathbf{V}\, \bb^{[1]}_{1}\, \mathbf{V}^{-1} &= \bb^{[1]}_{1} - \bb^{[2]\dagger}_{2}\bb^{[1]}_{2},\\
\mathbf{V}\, \bb^{[2]}_{2}\, \mathbf{V}^{-1} &= \bb^{[2]}_{2} - \bb^{[1]\dagger}_{1}\bb^{[1]}_{2},\\
\mathbf{V}\, \bb^{[1]\dagger}_{2}\, \mathbf{V}^{-1} &= \bb^{[1]\dagger}_{2} + \bb^{[1]\dagger}_{1}\bb^{[2]\dagger}_{2},
\end{split}
\end{equation}
respectively. Putting everything together, relabelling the oscillators,
\begin{equation}
\bb^{[1]}_{2} \to \bb^{[1]}_{1},\quad \bb^{[2]}_{1} \to \bb^{[1]}_{2},\quad
\bb^{[1]\dagger}_{2}\to \bb^{[1]\dagger}_{1},\quad \bb^{[2]\dagger}_{2} \to \bb^{[1]\dagger}_{2},
\end{equation}
and representing the generators of $\mathfrak{sl}(2)$ spins acting on $\mathcal{V}_{j}$ (where $2j=j_{1}-j_{2}+\lambda$) as
\begin{equation}
\begin{split}
\bJ^{11} &\leftarrow j_{1} + \lambda - \bb^{[1]\dagger}_{1}\bb^{[1]}_{1},\\
\bJ^{21} &\leftarrow \bb^{[1]\dagger}_{1}\bb^{[1]\dagger}_{1}\bb^{[1]}_{1} - (j_{1} - j_{2} - \lambda)\bb^{[1]\dagger}_{1},\\
\bJ^{12} &\leftarrow -\bb^{[1]},\\
\bJ^{22} &\leftarrow j_{2} - \bb^{[1]\dagger}_{1}\bb^{[1]}_{1},
\end{split}
\end{equation}
yields precisely the anticipated canonical representation of the mesonic Lax operator
\begin{equation}
\bL_{\{1,2\}}(z) =
\begin{pmatrix}
z + \bJ^{11} -\bh_{1} & \bJ^{21} - \bb^{\dagger}_{1}\bb_{2} & \bb^{\dagger}_{1} \\
\bJ^{12} - \bb^{\dagger}_{2}\bb_{1} & z + \bJ^{22} - \bh_{2} & \bb^{\dagger}_{2} \\
-\bb_{1} & -\bb_{2} & 1
\end{pmatrix}.
\end{equation}

\paragraph{$\mathfrak{gl}(1|1)$ case.}
It may be instructive to also explicitly spell out the fusion step \eqref{eqn:fusion_rule} for the $\mathfrak{gl}(1|1)$ Lie 
superalgebra. The latter is spanned by four elements, two bosonic generator $N$ and $E$, and two fermionic ones $\psi^{\pm}$.
Commutation relations read
\begin{equation}
\big[N,\psi^{\pm}\big]=\pm \psi^{\pm},\quad
\big[\psi^{-},\psi^{+}\big] = E,\quad
(\psi^{-})^{2}=(\psi^{+})^{2} = 0,
\label{eqn:gl11_abstract}
\end{equation}
where $E$ is the central element.
The defining (fundamental) representation is of dimension $2$, and is given by $2\times 2$ matrices
\begin{equation}
E = \begin{pmatrix}
1 & 0 \\
0 & 1
\end{pmatrix},\quad
N = \begin{pmatrix}
0 & 0 \\
0 & 1
\end{pmatrix},\quad
\psi = \begin{pmatrix}
0 & 1 \\
0 & 0
\end{pmatrix},\quad
\psi^{\dagger} = \begin{pmatrix}
0 & 0 \\
1 & 0
\end{pmatrix}.
\label{eqn:gl11_defining}
\end{equation}
By setting $E=0$, we obtain a trivial one-dimensional irreducible representation with $\psi^{\pm}\equiv 0$. There moreover exists
a family of two-dimensional irreducible representations, denoted by $\langle n,e \rangle$ (with $E\neq 0$), which reads
\begin{equation} 
E = \begin{pmatrix}
e & 0 \\
0 & e
\end{pmatrix},\quad
N = \begin{pmatrix}
n-1 & 0 \\
0 & n
\end{pmatrix},\quad
\psi = \begin{pmatrix}
0 & 1 \\
0 & 0
\end{pmatrix},\quad
\psi^{\dagger} = \begin{pmatrix}
0 & 0 \\
e & 0
\end{pmatrix},
\label{eqn:gl11_2dim}
\end{equation}
and includes the fundamental representation \eqref{eqn:gl11_defining} as $\langle 1,1 \rangle $.
Reducible indecomposable representations of $\mathfrak{gl}(1|1)$ (see e.g. \cite{RS92,SS06}) and not of our interest here.

Fusion in the fermionic case works as follows.
The partonic Lax operators contain a single fermionic specie and are of the form
\begin{equation}
\bL_{\{1\}}(z)  =
\begin{pmatrix}
z - (\bn-\tfrac{1}{2}) & \bc^{\dagger} \\
-\bc & 1
\end{pmatrix},\qquad
\bL_{\{2\}}(z) = 
\begin{pmatrix}
1 & \bc \\
\bc^{\dagger} & z + (\bn - \tfrac{1}{2})
\end{pmatrix}.
\end{equation}
By employing the universal fusion formula,
\begin{equation}
\bL^{[1]}_{\{1\}}(z-\tfrac{1}{2})\,\bL^{[2]}_{\{2\}}(z-\lambda+\tfrac{1}{2})=\bS\,\bL^{[1]}_{\lambda}(z)\,\bK^{[2]}\,\bS^{-1},
\end{equation}
we readily derive the $\mathfrak{gl}(1|1)$-invariant Lax operator which takes the form
\begin{equation}
\bL^{[1]}_{\lambda}(z) =
\begin{pmatrix}
z + \lambda - (\bn_{1}-\tfrac{1}{2}) & -2\lambda \bc^{\dagger}_{1} \\
-\bc_{1} & z - \lambda - (\bn_{1}-\tfrac{1}{2})
\end{pmatrix},
\end{equation}
where
\begin{equation}
\bK^{[2]} = 
\begin{pmatrix}
1 & \bc_{2} \\
0 & 1
\end{pmatrix},\qquad
\bS = \exp{\big(\bc^{\dagger}_{1} \bc^{\dagger}_{2}\big)},
\end{equation}
have been used. To match the canonical form of Eq.~\eqref{eqn:canonical_form}, given by
\begin{equation}
\bL_{\{1,2\}}(z) =
\begin{pmatrix}
z - \bJ^{11} & \bJ^{21} \\
-\bJ^{12} & z + \bJ^{22}
\end{pmatrix},
\end{equation}
the $\mathfrak{gl}(1|1)$ super spin generators $\bJ^{ab}$ should are identified as
\begin{align}
\bJ^{11} &= j_{1} + \bc^{\dagger}_{1}\bc_{1},&\quad
\bJ^{12} &= (j_{1}+j_{2}-\lambda)\bc^{\dagger}_{1},\\
\bJ^{21} &= \bc_{1},&\quad
\bJ^{22} &= j_{2} - \lambda - \bc^{\dagger}_{1}\bc_{1},
\end{align}
together with the following constraint on the $\mathfrak{gl}(1)$ scalars
\begin{equation}
j_{1}= -\tfrac{1}{2}-\lambda,\qquad j_{2}=\tfrac{1}{2}.
\end{equation}
Therefore, $\bL_{\lambda}(z)$ belongs to the two-dimensional representation $\mathcal{V}_{\lambda}$, with $\lambda$ being the 
central charge. A comparison with Eq.~\eqref{eqn:gl11_2dim} shows that $\mathcal{V}_{\lambda}\equiv \langle 1,2\lambda\rangle$.
In terms of fermionic algebra, operator $\bL_{\lambda}(z)$ admits an expansion
\begin{equation}
\bL_{\lambda}(z) = (-1)^{b}E^{ab}\stimes \bL^{ab}(\lambda)=z  + 2\lambda\,c\,\bc^{\dagger} + c^{\dagger}\,\bc - 2\lambda n - \bn.
\end{equation}
Recall that for $\lambda=0$, module $\mathcal{V}_{\lambda}$ becomes an atypical indecomposable representation (a short multiplet) 
which is no longer irreducible; it contains a one-dimensional invariant subspace corresponding to the Fock vacuum. However, these
exceptional instances do not seem to be relevant in the context of boundary-driven spin chains.

We moreover wish to emphasize that prescription \eqref{eqn:oscillator_realization} provides an explicit oscillator realization
of $\mathfrak{gl}(n|m)$ Lie superalgebras. Let us consider the $\mathfrak{gl}(2|1)$ case as an example, and fixing the
grading to $\bigotimes\!\!-\!\!\!-\!\!\!\bigotimes$. The bosonic subalgebra is a direct sum $\mathfrak{gl}(2)\oplus \mathfrak{u}(1)$, 
and is spanned by $\mathfrak{gl}(2)$ generators (writing $\bn_{i}=\bc^{\dagger}_{i}\bc_{i}$)
\begin{equation}
\begin{split}
\widehat{\bJ}^{11} = \bJ^{11}_{1} + \bn_{1},\qquad \widehat{\bJ}^{13} = \bJ^{13}_{1} + \bc^{\dagger}_{1}\bc_{2}, \\
\widehat{\bJ}^{33} = \bJ^{33}_{1} + \bn_{2},\qquad \widehat{\bJ}^{31} = \bJ^{31}_{1} + \bc^{\dagger}_{2}\bc_{1},
\end{split},
\end{equation}
and $\mathfrak{u}(1)$ generator
\begin{equation}
\widehat{\bJ}^{22} = j_{2} - \bn_{1} - \bn_{2}.
\end{equation}
In addition, there are four fermionic charges which are parametrized as
\begin{equation}
\begin{split}
\widehat{\bJ}^{12} = \bc^{\dagger}_{1}\widehat{\bJ}^{22} +
\bJ^{11}_{1}\bc^{\dagger}_{1} + \bJ^{13}_{1}\bc^{\dagger}_{2},\qquad \widehat{\bJ}^{21} = \bc_{1}, \\
\widehat{\bJ}^{32} = \bc^{\dagger}_{2}\widehat{\bJ}^{22} +
\bJ^{31}_{1}\bc^{\dagger}_{1} + \bJ^{33}_{1}\bc^{\dagger}_{2},\qquad \widehat{\bJ}^{23} = \bc_{2}.
\end{split}
\end{equation}
The $\mathfrak{sl}(1|2)$ case can be obtained by restricting $\bJ^{ab}_{1}$ to $\mathfrak{sl}(2)$ spins acting on 
$\mathcal{V}_{j_{1}}$, while $j_{2}$ is the remaining Dykin label.

\section{Non-interacting fermions}
\label{app:free}

In this section we provide the solution to the problem of boundary-driven non-interacting spinless fermions
hopping on a one-dimensional lattice. The Hamiltonian of the model can be seen as a Yang--Baxter integrable spin chain invariant
under $\mathfrak{gl}(1|1)$ Lie superalgebra.\footnote{The model can also be mapped to the XX Heisenberg spin chain Hamiltonian.
In the spin picture, we deal with a model invariant under the $q$-deformed quantum symmetry $\mathcal{U}_{q}(\mathfrak{sl}(2))$ for
the value of the deformation parameter $q=\ii$.}

In Section \ref{sec:fermionic} we constructed steady-state solutions for the simplest fermionic boundary reservoirs with
equal coupling strengths. Our aim here is to demonstrate that the problem of free fermions represents a special case which
even allows for solutions going beyond those discussed in Section \ref{sec:fermionic}.
Quite remarkably, the \emph{operator} Schmidt rank\footnote{The operator analogue of the Schmidt rank characterizes a degree of 
bipartite entanglement of a mixed quantum state. In the language of matrix-product states it coincides with the bond dimension.}
of $\rho_{\infty}$ now equals $4$ and thus does not depend on the system size.
This is in stark contrasted to the solutions pertaining to higher-rank symmetries which all exhibit Schmidt ranks 
which grow algebraically with system size. A finite Schmidt rank can be understood as a strong indication that the problem of finding 
the steady states may be tractable by `brute force', that is first by explicitly computing the null space of the Liouvillian generator 
$\mathcal{L}$ and subsequently analytically parametrizing the solution (e.g. with help of symbolic algebra routines).
An obvious advantage of this approach is that it does not require any prior knowledge of underlying algebraic structures.
This allows allows to conveniently parametrize the solutions directly in terms of physical couplings attributed 
to the boundary reservoirs.

We shall provide an extended four-parametric set of solutions for the asymmetric driving
\begin{equation}
A_{1} = \sqrt{g\,\zeta}\,\sigma^{+}_{1},\qquad A_{N} = \sqrt{g/\zeta}\,\sigma^{-},
\label{eqn:dissipators_ff}
\end{equation}
involving coupling rate parameters $g,\zeta \in \mathbb{R}$, supplemented with two additional boundary
external fields,
\begin{equation}
H_{\rm field} = \frac{h_{\rm L}}{2}\sigma^{z}_{1} + \frac{h_{\rm R}}{2}\sigma^{z}_{N},
\end{equation}
of unequal magnitudes $h_{\rm L},h_{\rm R} \in \mathbb{R}$.
We notice that the solutions given below appear to lie outside of $\mathfrak{gl}(1|1)$-invariant Lax operators of
Eqs.~\eqref{eqn:gl11_solution} and \eqref{eqn:gl11_solution_reversed}.\footnote{In practice it turns out that free fermions
allows even more general types of non-perturbative integrable boundaries which then those considered here.}

The $\bL$-operator for the $\Omega$-amplitude is now formally linked to a two-dimensional auxiliary representation denoted
by $\mathcal{V}_{\u}$. Here $\u$ is a four-component vector label which involves the boundary parameters,
$\u = (g,\zeta,h_{\rm L},h_{\rm R})$. In terms of Pauli matrices, the $\bL$-operator admits the expansion
\begin{align}
\bL_{\mathcal{V}_{\u}} &= \frac{\sigma^{0}}{2}\otimes
\begin{pmatrix}
\zeta+1 & 0 \\
0 & (\zeta-1)+(\tilde{h}_{\rm L}-\zeta\,\tilde{h}_{\rm R})
\end{pmatrix}
+ \frac{\sigma^{z}}{2}\otimes
\begin{pmatrix}
\zeta-1 & 0 \\
0 & \ii g(1-\zeta)-(\tilde{h}_{\rm L}+\zeta\,\tilde{h}_{\rm R})
\end{pmatrix} \nonumber \\
&+ \sigma^{-}\otimes
\begin{pmatrix}
0 & 0 \\
\ii g(\zeta^{2}+1)+\zeta\,\delta h & 0
\end{pmatrix}
+\sigma^{+}\otimes
\begin{pmatrix}
0 & 1 \\
0 & 0
\end{pmatrix},
\end{align}
using shorthand notations $\tilde{h}_{i}=h_{i}-1$ and $\delta h=h_{\rm L}-h_{\rm R}$.
It can easily be verified that the $\bL$-operator provides a \emph{multi-colored} family commuting of transfer matrices,
\begin{equation}
T(\u)={\rm Tr}_{\mathcal{V}_{\u}}\,\bL_{\mathcal{V}_{\u}}\otimes \cdots \otimes \bL_{\mathcal{V}_{\u}},
\qquad [T(\u_{1}),T(\u_2)] = 0
\qquad \forall \u_{1},\u_{2}\in \mathbb{C}^{4}.
\end{equation}
Involution property of $T(\u)$ is ensured by the multi-colored $6$-vertex $R$-matrix 
$\bR_{\mathcal{V}_{\u_{1}}\mathcal{V}_{\u_{2}}}$ which operates
on $\mathcal{V}_{\u_{1}}\otimes \mathcal{V}_{\u_{2}}$,
\begin{equation}
\bR_{\mathcal{V}_{\u_{1}}\mathcal{V}_{\u_{2}}} = 
\begin{pmatrix}
a_{1}(\u_{1},\u_{2}) & 0 & 0 & 0 \\
0 & b_{1}(\u_{1},\u_{2}) & c_{1}(\u_{1},\u_{2}) & 0 \\
0 & c_{2}(\u_{1},\u_{2}) & b_{2}(\u_{1},\u_{2}) & 0 \\
0 &0 & 0 & a_{2}(\u_{1},\u_{2})
\end{pmatrix},
\label{eqn:multicolor_R}
\end{equation}
intertwining two copies of auxiliary spaces $\mathcal{V}_{\u}$ associated to a pair of $\bL$-operators
$\bL_{\mathcal{V}_{\u_{i}}}$ acting on $\mathbb{C}^{2}\otimes \mathcal{V}_{\u_{i}}$,
\begin{equation}
\bR_{\mathcal{V}_{\u_{1}}\mathcal{V}_{\u_{2}}}\bL_{\mathcal{V}_{\u_{1}}}\bL_{\mathcal{V}_{\u_{2}}}=
\bL_{\mathcal{V}_{\u_{2}}}\bL_{\mathcal{V}_{\u_{1}}}\bR_{\mathcal{V}_{\u_{1}}\mathcal{V}_{\u_{2}}}.
\end{equation}
Moreover, it can easily be verified that the $\bR$-matrix embedded into three-fold tensor space obeys the Yang--Baxter equation
\begin{equation}
\bR_{\mathcal{V}_{\u_{1}}\mathcal{V}_{\u_{1}}}
\bR_{\mathcal{V}_{\u_{1}}\mathcal{V}_{\u_{3}}}
\bR_{\mathcal{V}_{\u_{2}}\mathcal{V}_{\u_{3}}}=
\bR_{\mathcal{V}_{\u_{2}}\mathcal{V}_{\u_{3}}}
\bR_{\mathcal{V}_{\u_{1}}\mathcal{V}_{\u_{3}}}
\bR_{\mathcal{V}_{\u_{1}}\mathcal{V}_{\u_{1}}}.
\end{equation}
The amplitudes (Boltzmann weights) of the $R$-matrix read explicitly
\begin{equation}
\begin{split}
a_{1}(\u_{1},\u_{2}) &= \zeta_{1} \zeta_{2}\,g_{1} + g_{2} - \ii\,\zeta_{2}(h_{\rm L,1}-h_{\rm R,2}) \\
a_{2}(\u_{1},\u_{2}) &= g_{1} + \zeta_{1}\zeta_{2} g_{2} - \ii\,\zeta_{1}(h_{\rm L,2}-h_{\rm R,1}), \\
b_{1}(\u_{1},\u_{2}) &= \zeta_{1}g_{2} - \zeta_{2}g_{1} - \ii\,\zeta_{1}\zeta_{2}(h_{\rm R,1}-h_{\rm R,2}), \\
b_{2}(\u_{1},\u_{2}) &= \zeta_{1}g_{1} - \zeta_{2}g_{2} - \ii(h_{\rm L,1}-h_{\rm L,2}), \\
c_{1}(\u_{1},\u_{2}) &= (1+\zeta^{2}_{2})g_{2} -\ii\,\alpha_{2}(h_{\rm L,2}-h_{\rm R,2}), \\
c_{2}(\u_{1},\u_{2}) &= (1+\zeta^{2}_{1})g_{1} -\ii\,\alpha_{1}(h_{\rm L,1}-h_{\rm R,1}),
\end{split}
\label{eqn:weight_ff}
\end{equation}
and satisfy the \emph{free fermion condition}~\cite{BS85},
\begin{equation}
a_{1}a_{2} + b_{1}b_{2} = c_{1}c_{2}.
\label{eqn:ff_condition}
\end{equation}
The differential Yang--Baxter relation on $\mathbb{C}^{2}\otimes \mathbb{C}^{2}$ yields the standard form of
the Sutherland equation,
\begin{equation}
\big[h^{1|1},\bL_{\mathcal{V}{\u}}\otimes \bL_{\mathcal{V}{\u}}\big] =
\bL_{\mathcal{V}_{\u}} \otimes \widetilde{\bL}_{\mathcal{V}_{\u}}-
\widetilde{\bL}_{\mathcal{V}_{\u}} \otimes \bL_{\mathcal{V}_{\u}},
\label{eqn:Sutherland_ff}
\end{equation}
where
\begin{equation}
\widetilde{\bL}_{\mathcal{V}_{\u}} = \tfrac{1}{2}(\ii\,g - \zeta h_{\rm R}) \sigma^{0} \otimes \boldsymbol{\sigma}^{0}
-\tfrac{1}{2}(\ii\,g \zeta + h_{\rm L}) \sigma^{z}\otimes \boldsymbol{\sigma}^{z}.
\end{equation}
By expanding it to the entire spin chain, we obtain the action of the unitary part $\hcalL_{0}$ which
in distinction to the canonical solutions discussed in the paper this time (assuming non-vanishing $\delta h$) acquires an
additional term,
\begin{equation}
[H,\Omega_{N}(\u)] = \tilde{g}
\begin{pmatrix}
\zeta\,\delta h + 1 & 0 \\
0 & \zeta
\end{pmatrix}\otimes \Omega_{N-1}(\u) + \tilde{g}\,\Omega_{N-1}(\u)\otimes
\begin{pmatrix}
1 & 0 \\
0 & -\zeta - \delta h
\end{pmatrix} + \delta h\,\Omega_{N}(\u).
\label{eqn:unitary_ff}
\end{equation}

\end{appendix}

\bibliography{SUSY.bib}

\nolinenumbers

\end{document}